\documentclass[12pt]{article}
\usepackage{graphicx} % Required for inserting images

\usepackage{geometry}
\usepackage{authblk}

\usepackage{parskip}

\usepackage[T1]{fontenc}

\usepackage[labelfont=bf]{caption}
\usepackage[labelsep=period]{caption}
\usepackage{booktabs}
\usepackage{tabularx} 
\usepackage{nicematrix}
\usepackage{multirow}
\usepackage{adjustbox}
\usepackage[usestackEOL]{stackengine}
\usepackage{floatrow} 
\usepackage{makecell}

\usepackage{xcolor}
\usepackage[colorlinks=true, urlcolor=blue,bookmarks=true,citecolor=blue]{hyperref}
\usepackage{color, colortbl}
\definecolor{Gray}{gray}{0.95}

\usepackage[sorting=none,
style=nature]{biblatex} 
\bibliography{bibliography} % Entries are in the refs.bib file

\usepackage{comment}
\usepackage{lscape}

\usepackage{paralist}

\usepackage{longtable}

\newcolumntype{Y}{>{\centering\arraybackslash}X}

\newcommand*{\belowrulesepcolor}[1]{% 
  \noalign{% 
    \kern-\belowrulesep 
    \begingroup 
      \color{#1}% 
      \hrule height\belowrulesep 
    \endgroup 
  }%
} 
\newcommand*{\aboverulesepcolor}[1]{% 
  \noalign{% 
    \begingroup 
      \color{#1}% 
      \hrule height\aboverulesep 
    \endgroup 
    \kern-\aboverulesep 
  }%
}

\usepackage{cellspace} 
\usepackage{amsmath}

\title{Clinical translation of machine learning algorithms for seizure detection in scalp electroencephalography: systematic review}
\author[1]{Nina Moutonnet}
\author[3]{Steven White}
\author[4]{Benjamin P Campbell}
\author[5]{Saeid Sanei}
\author[6]{Toshihisa Tanaka}
\author[7]{Hong Ji}
\author[5,8]{Danilo Mandic}
\author[2,8]{Gregory Scott}
\affil[1]{Department of Computing, Imperial College London}
\affil[2]{Department of Brain Sciences, Imperial College London}
\affil[3]{Department of Clinical Neurophysiology, National Hospital for Neurology \& Neurosurgery, London, United Kingdom}
\affil[4]{Department of Bioengineering, Imperial College London}
\affil[5]{Department of Electrical and Electronic Engineering, Imperial College London}
\affil[6]{Tokyo University of Agriculture and Technology}
\affil[7]{Shaanxi Provincial Key Laboratory of Fashion Design Intelligence, Xi'an Polytechnic University}
\affil[8]{UK Dementia Research Institute}

\date{}

\begin{document}

\setlength{\parindent}{0pt}

\maketitle
\section*{Abstract}
% Abstract to be embedded in main document
Machine learning algorithms for seizure detection have shown considerable diagnostic potential, with recent reported accuracies reaching 100\%. Yet, only few published algorithms have fully addressed the requirements for successful clinical translation. This is, for example, because the properties of training data may limit the generalisability of algorithms, algorithm performance may vary depending on which electroencephalogram (EEG) acquisition hardware was used, or run-time processing costs may be prohibitive to real-time clinical use cases. To address these issues in a critical manner, we systematically review machine learning algorithms for seizure detection with a focus on clinical translatability, assessed by criteria including generalisability, run-time costs, explainability, and clinically-relevant performance metrics. For non-specialists, the domain-specific knowledge necessary to contextualise model development and evaluation is provided. It is our hope that such critical evaluation of machine learning algorithms with respect to their potential real-world effectiveness can help accelerate clinical translation and identify gaps in the current seizure detection literature. 

\section*{Corresponding author}
Dr Gregory Scott BEng MSc MBBS MRCP PhD
Post-Doctoral, Post-CCT Research Fellow 
Honorary Consultant Neurologist
UK DRI Care Research and Technology Centre
9th Floor, Sir Michael Uren Hub,
Imperial College London, 86 Wood Lane,
London. W12 0BZ. UK.
Tel: +44 (0)7909 691484
Email: gregory.scott99@imperial.ac.uk

\section*{Keywords} clinical translation; deep learning; electroencephalography; epilepsy; machine learning; scalp; seizure detection 
\newpage

\section{Introduction}
A seizure is an abnormal synchronous excitation of one or more populations of neurons in the brain. Seizures are not a rare phenomenon, with an average lifetime incidence of 2-5\% \cite{neligan2009incidence}. Seizures may occur in a range of clinical contexts, in particular in patients with \emph{epilepsy} - defined as a tendency to unprovoked seizures \cite{fisher2014ilae}. Other, provoking, causes of seizures include central nervous system (CNS) infections, metabolic abnormalities, traumatic brain injuries, and drug toxicity. Seizure symptoms vary widely, from abnormal sensations to convulsions and altered awareness. Seizures are usually self-limiting, in that they typically last less than two minutes. Occasionally, they may continue for more than five minutes and/or recur without full recovery, a state termed \emph{status epilepticus}, a medical emergency \cite{falco2016treatment}.  

The accurate and timely detection of seizures is an important healthcare challenge \cite{elger2018diagnostic}. Seizure detection is straightforward when the seizure activity has a well-recognised clinical correlate like generalised convulsions or a subjective alteration in experience, termed an \emph{aura}. However, seizures without clear symptoms may be missed or mistaken for other medical phenomena. For example, ongoing seizure activity without an obvious motor component, known as \emph{non-convulsive status epilepticus} (NCSE), affects 8-20\% of patients in intensive care units (ICUs) and can be fatal \cite{laccheo2015non}. Despite being a medical emergency, it is critically underdiagnosed because it can go unnoticed \cite{kaplan1999assessing}. Patients with established epilepsy frequently under-report their seizures, as they may occur during sleep, or impair patients' awareness and memory of the events \cite{elger2018diagnostic}. In contrast, patients with \textit{non-epileptic attacks} (NEAs) - events appearing superficially similar to seizures but without the associated abnormal brain activity - are at risk of misdiagnosis and inappropriate treatment \cite{jungilligens2021misdiagnosis}.  

This highlights the need for an automated and accurate method for seizure detection. Among biological signals, accelerometry, electromyography (EMG), and electroencephalography (EEG) are most frequently used for seizure detection
\cite{naganur2022automated, elger2018diagnostic}. Of these methods, scalp EEG is most informative, as it contains subtle electrical abnormalities even in the absence of clinical correlates \cite{sanei2021eeg}. For this reason, it is the focus of this work.

\begin{figure}[h]
    \includegraphics[width=\textwidth]{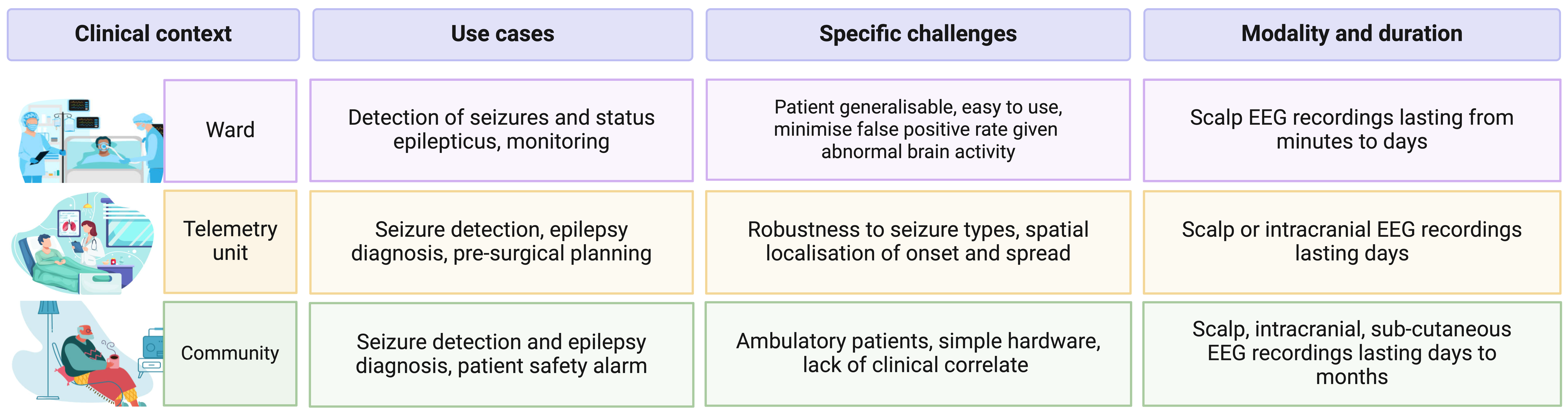}
    \caption{Potential use cases, challenges and recording modalities for automated scalp EEG seizure detection (Created with BioRender, \cite{created_with_biorender}). Applications of seizure detection algorithms range widely, from (1) highlighting to clinicians sections of interest in long recordings to simplify offline annotation; (2) real-time seizure detection using continuous EEG in ambulatory patients, telemetry units, or ICU; to (3) automatic recording of seizure diary entries.}
    \label{fig:usecases}
\end{figure}

The EEG recorded from the scalp is used to non-invasively measure the electrical activity of the brain. Traditionally, scalp EEG traces are examined by a neurophysiologist to rate seizure-related disorders \cite{smith2005eeg}. However, resource constraints as well as significant inter-rater variability currently limit both the availability and quality of EEG \cite{patel2015management, gerber2008interobserver, ronner2009inter}. A health economics analysis concluded that continuous EEG monitoring accounted for an average of 5\% of total hospital charges for patients monitored in intensive care \cite{ney2013continuous}. In practice, EEG monitoring is rationed due to the scarcity of experts capable of interpreting it \cite{abend2015much, patel2015management, Uk_parliament_health_com}. It is therefore envisaged that in the future of e-Health, together with wearable sensors \cite{mandic2023your}, automated EEG-based seizure detection algorithms will improve the availability of EEG and the care of patients with seizures.

Despite successes, several challenges for automated seizure detection from scalp EEG remain. We have identified three main challenges which remain obstacles towards a widespread adoption of automated seizure detection:
\begin{itemize}
    \item Data complexity,
    \item Seizure definition,
    \item Data collection and labelling discrepancies.
\end{itemize}

\textbf{Data complexity}: Scalp EEG data is multi-dimensional, non-stationary, imbalanced, and typically corrupted by artefacts (e.g. muscle activity, eye movements). Combining such complex data with the multiple manifestations and symptoms of seizures results in high variability of seizure EEG recordings. Without consideration of this variability, an algorithm designed for one seizure type or recording centre may fail if applied in a different centre or to a different patient population. Furthermore, seizures with a deep brain origin, such as the mesial temporal lobes, may be undetectable by standard interpretation of scalp EEG (so-called \textit{scalp-negative} seizures \cite{lam2016widespread}). 

\textbf{Seizure definition}: There is no single benchmark for identifying a seizure. The nosology of epilepsy and seizures is an active area of debate in epileptology, with many different systems of classification  \cite{scheffer2017ilae, luders1998semiological, luders2019critique}. This is particularly relevant for the definition of \textit{status epilepticus} and its subtypes \cite{trinka2015eeg, luders2019critique}. Another complication is the possibility of seizures being described as a non-binary phenomenon \cite{kalamangalam2018ictal}. 

\textbf{Data collection and labelling discrepancies}: In manual clinical EEG interpretation and labelling practice there are significant discrepancies between what experts consider to be seizure activity. This results in labeling differences, in particular regarding the timings of seizure onset and offset \cite{ronner2009inter, grant2014eeg, scheuer2021seizure}. There are a range of focal non-seizure related waveform abnormalities which may be observed clinically, where some may be mistaken for true seizures, while the clinical value of others is still debated \cite{foreman2012generalized, gelisse2021lateralized}. Other sources of variability arise from differences in EEG hardware and recording configurations.

The utility of machine learning (ML) for seizure detection has shown considerable promise, with reported accuracies reaching 100\% \cite{abdelhameed2021deep}. However, published ML algorithms typically do not fully address the three key challenges outlined above, which impacts their generalisability and utility. Furthermore, patient-specific and cross-patient use cases are frequently conflated. Many algorithms are trained and evaluated only on a single patient's data (e.g. \cite{usman2021detection, saidi2021novel, cura2020epileptic, das2023novel}). While patient-specific approaches may achieve higher classification performance compared to cross-patient algorithms, real-world use cases for patient-specific seizure detection algorithms are far more limited than for algorithms that can generalise across patients (see Figure \ref{fig:usecases}). For instance, patient-specific algorithms are well suited to long-term ambulatory seizure monitoring, but an algorithm used in the ICU or A\&E should be capable of generalising across patients. 

\subsection{Structure of this review}

To keep this article self-contained, we start from the necessary domain specific background about EEG recordings and seizures, which is suitable for non-specialists (Section \ref{domain_section}). Next, the key properties of publicly available datasets for seizure detection are described (Section \ref{dataset_section}). This is followed by an outline of the preferred reporting items for systematic reviews and meta-analyses (PRISMA) guidelines used to select articles (Section \ref{prisma_section}). We then comprehensively review the selected automated seizure detection algorithms, ramping from data pre-processing through to performance metrics (Section \ref{review_section}). Finally, we highlight potential research gaps and challenges (Section \ref{question_section}), and outline guidelines for the development of future EEG-based seizure detection algorithms, especially in the context of e-Health and care in the community (Section \ref{guidelines}).

\newpage
\section{Domain-specific knowledge} \label{domain_section}
\subsection{EEG recording techniques}\label{EEG_rec_techniques}

The scale and complexity of neuronal activity make microscopic brain dynamics difficult to assess. However, the currents generated by populations of neurons have an amplitude sufficiently high to measure voltages related to brain activity using EEG electrodes \cite{teplan2002fundamentals}. The EEG electrodes can be placed on the scalp, or implanted in the brain or below the skin. These two kinds of EEG recording arrangements are known as \emph{extracranial} and \emph{intracranial} techniques, respectively. Intracranial EEG can be recorded using methods such as \emph{electrocorticography} (ECoG)  or \emph{stereotaxic EEG}. The former uses strips or grids of ~50-100 electrodes implanted over an area of cortex via a \emph{craniotomy} procedure to produce a skull window \cite{lesser2010subdural}.  The latter uses ~5-15 wired electrodes penetrating the brain via multiple small \emph{burr holes} in the skull \cite{parvizi2018human}, targeting deeper structures like the hippocampus.

Intracranial EEG is advantageous in terms of signal quality because electrodes can be placed in the brain regions of interest, whereas activity recorded extracranially consists of a superposition of various brain sources. Additionally, intracranial electrodes generate a much higher signal-to-noise ratio than scalp electrodes, and the recorded signal is less likely to be corrupted by artefacts (e.g. muscle activity) \cite{shorvon2012oxford}. Therefore, algorithms trained on intracranial recording typically provide more accurate seizure detection than those trained on extracranial recordings \cite{casale2022sensitivity}. However, fitting intracranial electrodes comes with considerable medical risk and is costly. This makes intracranial EEG recording relatively rare and generally confined to epilepsy surgery planning  (see Figure \ref{fig:usecases}). The locations of intracranial EEG electrodes are tailored to an individual patient’s case rather than defined by a standardised system, and since most patients have seizures originating from temporal or frontal lobes, it is rare to find published intracranial EEG data from non-temporal and non-frontal sites. For these reasons, we do not address ML algorithms for intracranial EEG in this review.

Another recording technique not discussed in this review is minimally invasive \emph{sub-cutaneous EEG}, as it is still not widely established. This has recently been developed to provide comparable signal quality to scalp EEG and allow high duration continuous recordings in ambulatory patients with seizures  \cite{viana2021230, duun2015eeg}. 

Scalp EEG is by far the most common clinical technique for recording brain activity in seizure-related disorders. In this review, we focus on scalp EEG recordings and ML algorithms developed for this modality.

\subsection{Scalp EEG} \label{sEEG_section}

\begin{figure}[h]
        \caption{The 10-20 electrode placement system with front-back (nasion to inion) 10\% and 20\% electrode distances. The spatial resolution of scalp EEG setups can range from as few as 1 channel (low resolution) to 256 channels (high resolution). Each EEG electrode is labelled by one or two alphabetic characters followed by a digit. This combination describes the location of each electrode. The alphabetic characters are Fp for frontal-polar, F for frontal, P for parietal, T for temporal, O for occipital, and C for the central region of the brain. Odd-numbered electrodes are located on the left side of the brain, even-numbered electrodes on the right side, and Z electrodes lie along the midline of the scalp. \cite{shriram2013eeg}  (Created with BioRender, \cite{created_with_biorender_1020})}
        \includegraphics[width=\textwidth]{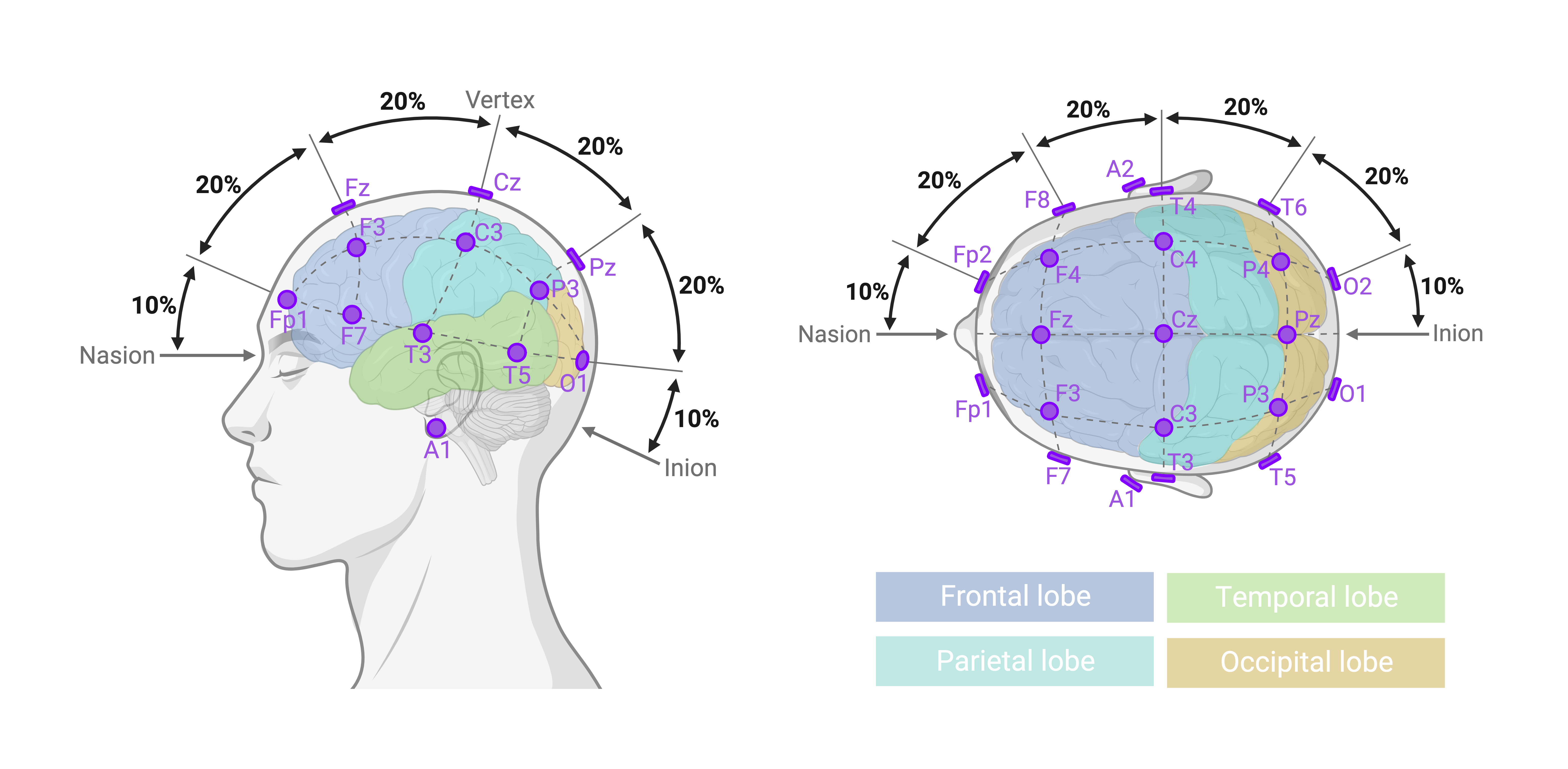}
        \label{fig:tentwenty}
\end{figure}

Scalp EEG typically uses about 20 electrodes affixed across the skin in standardised locations typically using a cap equipped with Ag-AgCl electrodes. A widely used international system of electrode placement and naming is the \emph{10-20 system}, which refers to the distances between adjacent electrodes being either 10\% or 20\% of the total distance across the skull, based on anatomical landmarks (see Figure \ref{fig:tentwenty}). A deficiency of the 10-20 system is the inadequate coverage of the lower part of the brain, below the circumferential line formed by the electrodes Fp2, F8, T4, T6, 02, O1, T5, T3, F7, Fp1. The implication of this lack of coverage was evaluated in 1982 by \citeauthor{binnie1982practical} \cite{binnie1982practical}. They recorded fronto-temporal seizure activity in 100 patients, adding two anterior temporal electrodes to the 10-20 system, located below the standard F7 and F8 electrodes respectively. They found that of the 98 identified epileptiform abnormalities, 43 were maximally recorded by one of the two anterior temporal electrodes. This confirms that the 10-20 system is sub-optimal, and that the lowest circumferential row of electrodes should be placed lower to enhance the coverage of brain activity. 

Scalp EEG activity is primarily generated by cortical pyramidal neurons that are oriented perpendicularly to the surface of the brain. The amplitude of EEG is about  10-100 µV when measured on the scalp, which roughly is an order of magnitude lower than when measured intracranially. Clinical EEG is usually sampled at a frequency of 256–512 Hz.

\begin{figure}[H]
    \centering
    \includegraphics[width=\textwidth]{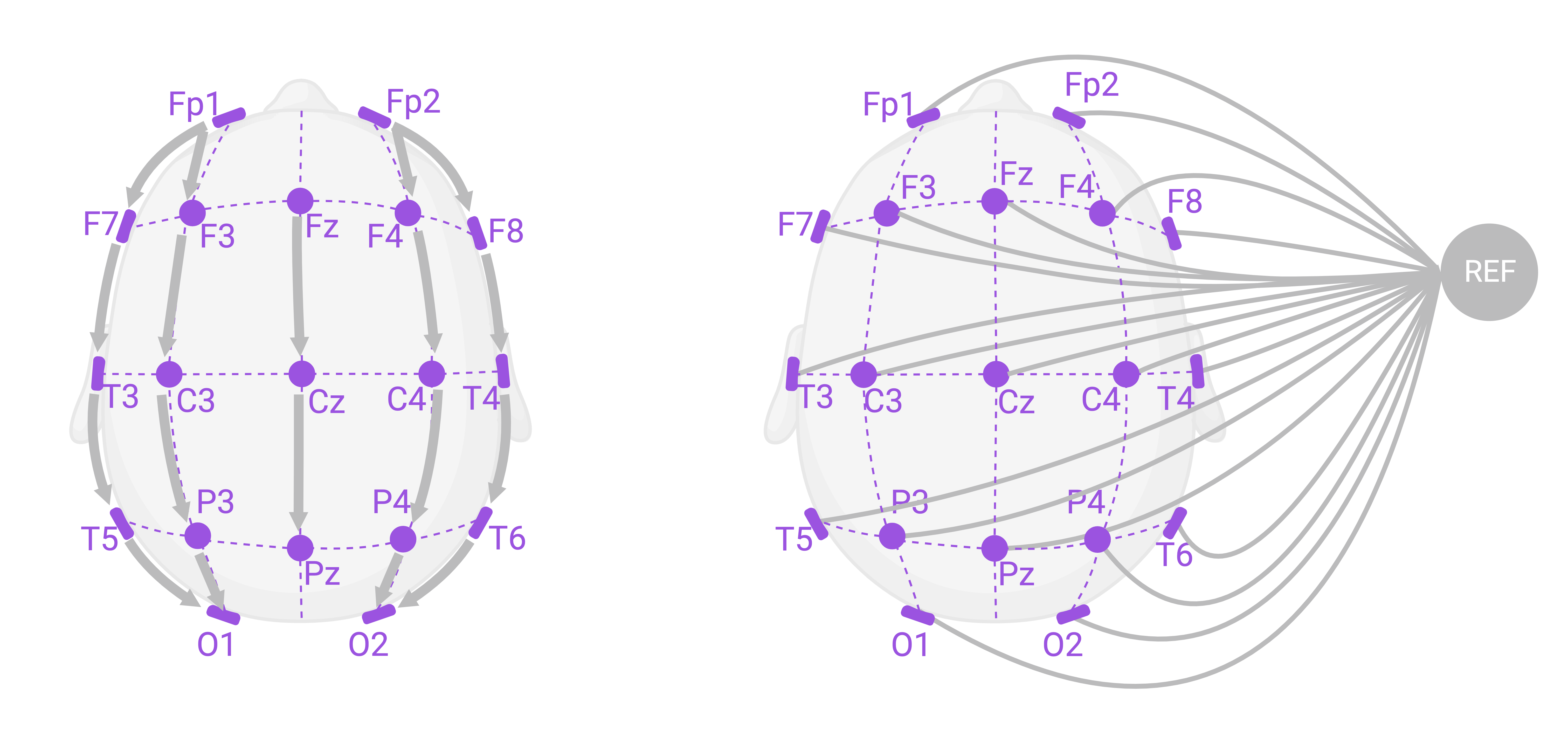}
    \caption{Bipolar and monopolar EEG montage. (Left) Double banana bipolar montage, where each electrode is referenced to the electrode behind it. Observe an outside temporal chain (e.g. Fp2-F8-T4-T6-O2) and an inside parasagittal chain (e.g. Fp2-F4-C4-P4-O2) on each side of the scalp, and a unique central chain in the middle (Fz-Cz-Pz). (Right) Monopolar montage, where all the electrodes are referenced to a single point. (Created with BioRender, \cite{created_with_biorender_1020})}
    \label{fig:eeg montage}
\end{figure}

The term EEG montage refers to a specific arrangement of scalp electrodes as well as how the electrodes are connected to one another. In a \textit{monopolar montage}, each electrode is connected to a common reference. This reference electrode can be located on the scalp or body, an example is an auricular (ear) electrode. Alternatively, the reference can be obtained by averaging the recorded potential of all electrodes, called \textit{average reference}. \cite{hu2018reference, niedermeyer2005electroencephalography}. In a \textit{bipolar montage}, each electrode is connected to one or two neighbouring electrodes \cite{britton2016electroencephalography} (see Figure \ref{fig:eeg montage}). It is sometimes possible to re-reference the data post recording with simple arithmetic steps which can be useful to bring new dynamics of interest to light.

The duration of EEG recordings for the diagnosis and management of seizure-related disorders varies from minutes to days (see Figure \ref{fig:usecases}). In practice, EEG recordings are often combined with video footage of the patient to provide information about their clinical status and behaviour, although video is not often available in public datasets. Video-EEG monitoring is typically used for diagnostic clarification of seizure-related disorders and for pre-surgical planning purposes \cite{navarro2022lessons}.

The context of EEG recording not only affects the inclusion of video data but also the duration and technical properties of the EEG (see Figure \ref{fig:usecases}). This includes variables such as the number of channels, the likelihood of seizure occurrence, and the nature of artefacts. For example, interference of muscle activity may corrupt the EEG of freely-moving awake patients more than that of motionless and comatose patients. The majority of clinical EEG recordings are acquired in a hospital, either on in-patients (often on an intensive care unit), patients undergoing video-EEG monitoring (in a video telemetry unit), or out-patients (a \emph{routine} EEG). Rarely is seizure-related EEG recorded in the community by a portable \emph{ambulatory} EEG device \cite{lawley2015role}.

\subsection{Conventional EEG frequency bands} 

The conventional frequency bands in EEG are specific ranges of frequencies used to categorise the electrical activity of the brain. The five frequency bands widely used are: \(\delta\) (0.5–4 Hz), \(\theta\) (4–8 Hz), \(\alpha\) (8–13 Hz), \(\beta\) (13–30 Hz), and \(\gamma\) (30–80 Hz).  These bands are associated with different cognitive and physiological states. For example, increased power in the \(\delta\) band is indicative of deep sleep; higher \(\theta\) band power is linked to states of relaxation;  \(\alpha\)\ band power is more prominent in eyes-closed restfulness; high \(\beta\) band power is associated with mental effort; and, increased \(\gamma\) band power has been linked to perception and conscious awareness \cite{bajaj2020wavelets}. Supporting our claim about the uncertainty in all aspects of EEG analyses, there is no consensus in the range of frequencies of these bands. For instance, while most sources define \(\gamma\) as 30–80 Hz, some define it as 30–100 Hz or even (30–200 Hz). 

\subsection{Seizure manifestations in scalp EEG}
Domain knowledge of the different manifestations of seizures is a prerequisite for anyone developing ML algorithms for seizure detection as failure to incorporate this knowledge is likely to the limit the robustness and generalisability of the developed algorithms. 

\subsubsection{Seizure types, duration, and evolution}
\begin{figure}[H]
        \caption{Main seizure types and some EEG characteristics. (Top) Normal brain activity, focal seizure and focal onset seizure with secondary generalisation alongside their EEG correlate. (Bottom) Nomenclature of seizure phases including demonstrative EEG segments of inter-ictal, pre-ictal, ictal and post-ictal activity. (Created with BioRender, \cite{created_with_biorender_seizure})}
        
        \includegraphics[width=\textwidth]{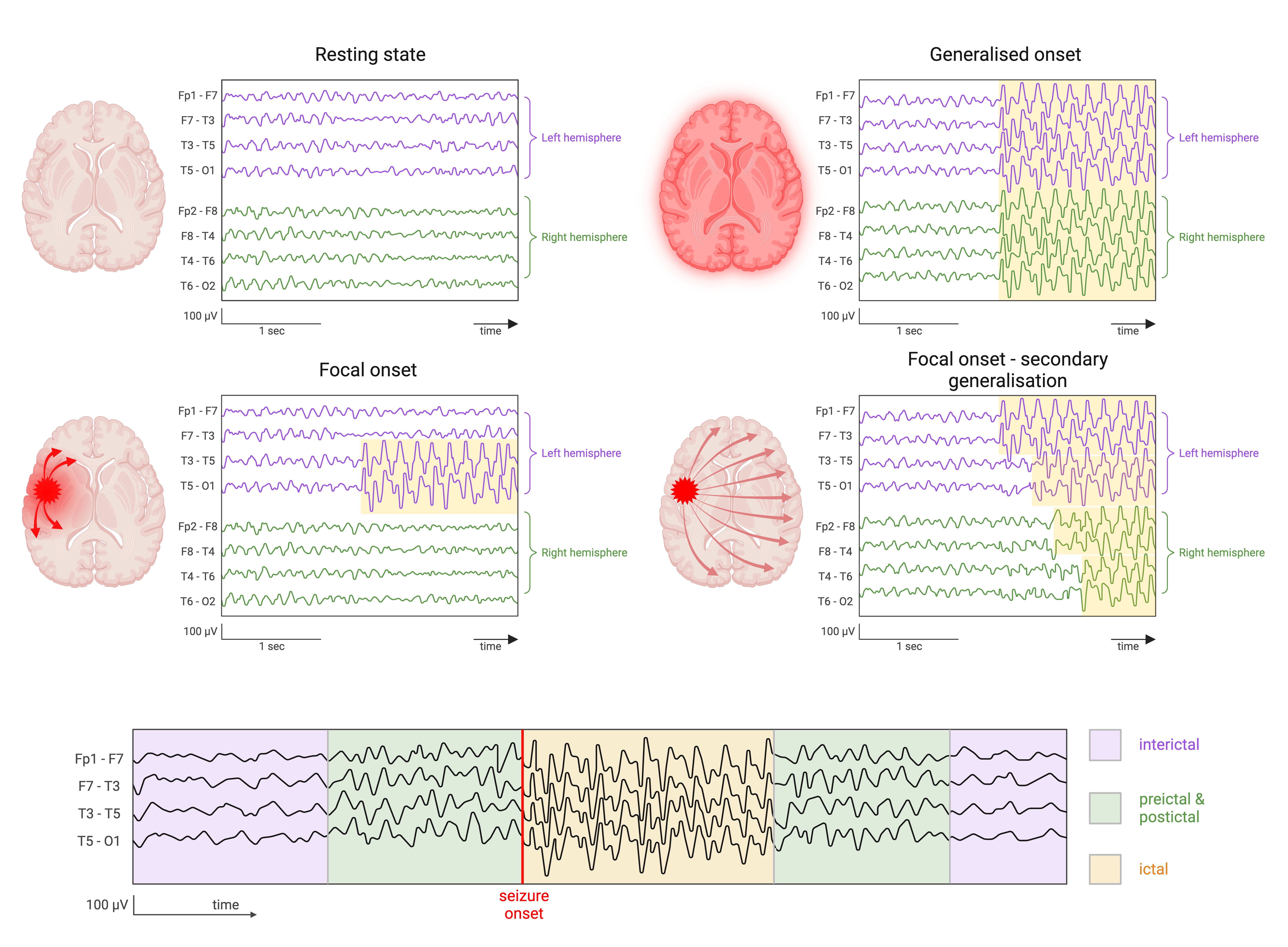}
        \label{fig:eegeg}
\end{figure}

A seizure is characterised by abnormal and excessive electrical activity from one or more populations of cortical neurons. Seizures are broadly classified into two categories: \emph{focal} (or partial) seizures and \emph{generalised} seizures. Focal seizures involve specific cortical region(s) confined to one hemisphere, while generalised seizures involve widespread electrical activity involving both hemispheres (see Figure \ref{fig:eegeg}). 

\textbf{Seizures vary in duration.} Most seizures last from a few seconds to two minutes, and are self limiting. The rate of seizures within an individual varies widely (e.g. from <1 seizure/year to >5/day) \cite{meritam2023duration}. Rarely, seizures continue for more than five minutes or occur recurrently without recovery of normal consciousness in between, termed \textit{status epilepticus. }

\textbf{A seizure can spatially evolve over time.} The starting location is known as the \emph{seizure onset zone} (\emph{SOZ}). A \emph{focal-onset} seizure is one that begins in specific brain region(s) in one hemisphere. A \emph{generalised-onset} seizure begins with widespread activity in both hemispheres. A focal-onset seizure with \emph{secondary generalisation} spreads to both hemispheres (see Figure \ref{fig:eegeg}). The spread of a seizure can happen on the order of seconds or minutes. Typical patterns of the spatial spread is outlined in \cite{jirsa2014nature}.

\subsubsection{Seizure-related phases} \label{Seizure-related_phases}
Every seizure has four periods or phases, which have distinct clinical and electrographic properties (see Figure \ref{fig:eegeg}) \cite{fisher2014can}: 

(1) \emph{Inter-ictal} - the period absent from any seizure, when there is no seizure activity and the patient and EEG are at \emph{baseline}.  This baseline may contain abnormalities which depend on the underlying seizure disorder and clinical status of the patient. For instance, the EEG of individuals with epilepsy may contain \emph{inter-ictal epileptiform discharges} (\emph{IEDs}) that can occur several times per minute \cite{kural2020criteria}. See \cite{abdi2023review} for a detailed review of signal processing and ML techniques for IED detection.  

(2) \emph{Pre-ictal} - the period immediately before a seizure, where the EEG may not be at \emph{baseline}. This phase is particularly relevant for EEG-based seizure \textit{prediction} algorithms (as opposed to \textit{detection}) that aim to anticipate the subsequent ictal phase \cite{corsini2006epileptic}. In this period, a patient may experience forewarning symptoms of an impending seizure.

(3) \emph{Ictal} - the period of seizure activity. The symptoms, and EEG characteristics depend on the brain region(s) affected by seizure activity. Potential spreading of the seizure activity occurs during this phase, causing changes in symptoms and EEG characteristics \cite{kural2020criteria}.

(4) \emph{Post-ictal} - the period immediately after a seizure, typically lasting seconds to minutes. The patient may be confused or drowsy, with EEG abnormalities distinct from those in the pre-ictal and ictal phases \cite{pottkamper2020postictal}.

Despite seizures having four phases, the transition between phases is not always sudden or clearly defined. This poses challenges for the development of ML algorithms for seizure detection \cite{tao2020ictal}.

\subsubsection{Epileptiform activity}
\emph{Epileptiform activity} refers to EEG patterns commonly associated with seizures and underlying seizure disorders, as illustrated in Figure \ref{fig:eegeg}. 

The distribution of epileptiform activity across EEG channels provides information about the spatial localisation of underlying brain activity. In generalised-onset or secondary generalised seizures, epileptiform activity is present in channels covering both hemispheres. Typically, all channels display epileptiform activity in the form of generalised spike-and-wave or polyspike-and-wave patterns (see Figure \ref{fig:eegeg}). 

Whilst epileptiform activity is a hallmark of the ictal phase of a seizure, it is not pathognomic. Epileptiform activity can be seen during any phase of an EEG recording, including the inter-ictal phase. Seizure detection algorithm designers should therefore not assume that the presence/absence of epileptiform activity constitutes a perfect seizure/non-seizure distinction. 

Similarly, the \textit{background} or \textit{baseline} EEG of a patient not having a seizure can sometimes display highly-abnormal looking EEG with non-epileptiform focal patterns. Those patterns may share some of the characteristics of epileptiform activity. Examples include periodic lateralised epileptiform activity (PLEDs), now commonly referred to as \textit{lateralised periodic discharges} (LPDs); and generalised rhythmic delta activity (GRDA). However, despite the patient not displaying typical seizure symptoms, the clinical significance of these patterns is still debated  \cite{hughes2010periodic}. Some argue that there exists somewhat arbitrarily-defined cut-offs for what constitutes true ictal activity  \cite{tao2020ictal}. For algorithm designers, the unfortunate - but important - takeaways are (1) that not all focal abnormalities constitute epileptiform activity, but some do, (2) not all epileptiform activity constitutes seizure activity.

\subsubsection{Seizure subtypes}
Beyond the focal/generalised distinction, seizures can be subdivided according to clinical, behavioral and electrographic manifestations. A number of different classification systems exist (e.g. \cite{scheffer2017ilae, luders1998semiological}).

In the publicly available Temple University Hospital (TUH) seizure corpus, the seizure subtype labels and their definitions, from most- to least- frequent, are: \emph{focal seizure} (FNSZ), typical duration 10s-2 min, with or without altered awareness, and with or without secondary generalisation; \emph{simple partial seizure} (SPSZ), a focal seizure without loss of awareness; \emph{complex partial seizure} (CPSZ), a focal seizure with loss of awareness; \emph{generalised tonic-clonic seizure} (TCSZ), a generalised seizure with convulsions; \emph{absence seizure} (ABSZ), a generalised seizure with altered awareness but no convulsions. Rarer subtypes include \emph{myoclonic seizure} (MYSZ) and \emph{tonic seizure} (TNSZ). The acronyms used above reproduce the naming convention used in the TUH seizure corpus.

Most datasets however, do not include labelling of seizure subtypes. This potentially limits the clinical validity of seizure detection algorithms as they might be applied to data where the distribution of seizure subtype differs from that of the training data. For example, absence seizures and tonic seizures are most frequently observed in children with genetic epilepsies, whereas the most common type of seizure in a hospitalised adult is a focal seizure with or without secondary generalisation. Ensuring that the algorithm is capable of detecting both type of seizure activity is critical. 

\subsubsection{Non convulsive status epilepticus}\label{NCSE_section}
An important use-case for clinical EEG is diagnosing \emph{status epilepticus} (previously defined), particularly non-convulsive status epilepticus (NCSE). This  can be challenging to diagnose because the seizure activity often lacks an observable correlate, such as convulsions. The lack of continuous EEG recording and concurrent interpretation is the reason for the underdiagnosis of NCSE.

Epileptiform activity for NCSE can be different to the epileptiform activity observed during `self-limiting' seizures \cite{trinka2015eeg}. Furthermore, the EEG abnormality during a seizure,  particularly during the protracted duration of status epilepticus (from minutes to hours), can change over time. Consequently, a classifier trained on data from isolated seizures may not accurately detect status epilepticus. 

\subsubsection{Scalp EEG corrupting artefacts}
Scalp EEG is prone to a range of recording artefacts. Whilst their impact can be reduced using EEG pre-processing techniques \cite{jiang2019removal}, an awareness of the characteristics of EEG artefacts and their relation to seizures is important for the development of seizure detection algorithms.

\begin{itemize}
    \item Electrical and environmental interference: External sources of electrical interference, such as nearby electronic devices or strong electromagnetic fields (e.g. 50Hz or 60Hz power line noise) can affect the EEG recording.
    
    \item Ocular artefacts: These result from eye movements and blinks, producing an electrical potential around the eyes which spreads across the scalp and contaminates the EEG recordings, particularly in frontal electrodes. The amplitude of blinking artefacts is generally much larger (in the order of hundreds of microvolts) compared to that of background EEG activity (a few to tens of microvolts).
    
    \item Muscle artefacts: These result from electrical activity and motion associated with muscle contraction. This includes eye movement, jaw clenching, facial grimacing, and large movements of the body such as convulsions during a generalised tonic-clonic seizure. Muscle artefacts are more prevalent in awake patients and often appear as high-amplitude, high-frequency signals.
    
    \item Cardiac and respiratory artefacts: Electrocardiographic (ECG) artefacts can be detected in some EEG channels. Despite considerable distance between the head and the heart, these artefacts can appear as regular spikes or waves in the recording. Breathing related artefacts can occur due to chest movement or changes in scalp impedance during respiration. These artefacts often appear as rhythmic fluctuations in the EEG.
    
    \item Other: The conductive properties of the gel placed between the skin and the electrode deteriorates over time, increasing the impedance between the electrode and the skin and resulting in signals with a lower SNR.
\end{itemize}

The impact of artefacts varies according to the clinical context (e.g. out-patient facilities versus intensive care units), the recording equipment and channels used, and the clinical status of the patient. Similarly, the artefacts can vary during the different phases of a seizure. In particular, motion artefacts, such as convulsions, can predominate the measured brain activity in EEG recordings occurring during a seizure. In fact, it is possible that a seizure detection algorithm may inadvertently be trained to detect correlated artefacts rather than the presence of genuine ictal activity.

\newpage
\section{Datasets} \label{dataset_section}
In this section, we highlight important properties of a range of available datasets for seizure detection applications. We focus exclusively on publicly accessible datasets containing scalp EEG recordings of ictal and non-ictal activity. The majority of published seizure detection algorithms have been trained on one or more of these datasets. However, some algorithms have been developed using proprietary scalp datasets or exclusively intracranial datasets (e.g. \cite{fu2020automatic, lyu2021automatic, song2020feature}). These have been omitted from this review as the evaluation of their clinical implementation is far more limited. For publicly available intracranial EEG datasets, see \cite{andrzejak2001indications, KAROLY2018, Kaggle_Mayo_EEG_data, AES_Seizure_Prediction_Challenge, MU_AES_Seizure_Prediction}, reviewed in \cite{wong2023eeg}. 

At the time of writing, there are six relevant datasets: the Children’s Hospital Boston Massachusetts Institute of Technology scalp EEG database \cite{shoeb2009application} (CHB-MIT), the Neurology and sleep centre Hauz Khas database (NSC-HK) \cite{swami2016eeg}, the TUH EEG seizure corpus (TUSZ) \cite{obeid2016temple}, the Helsinki University Hospital EEG database (HUH) \cite{stevenson2019dataset}, the Siena scalp EEG database (SSE) \cite{detti2020eeg}, and the American University of Beirut medical center dataset \cite{mendeleyDataBeirut}. Tables \ref{table1} and \ref{table3} summarise the dataset size and patient demographics; and the basic EEG data characteristics, respectively. 

\begin{table}[H]
    \footnotesize
    \centering
    \begin{NiceTabularX}{\textwidth}{Ycccc}
        \toprule
        \textbf{Dataset} & \textbf{Acronym} & \textbf{\makecell{Publication\\date}} & \makecell{\textbf{Number of} \\ \textbf{subjects}} & \textbf{Population} \\
        \midrule
        \belowrulesepcolor{Gray}
        \rowcolor{Gray}
        
        \href{https://isip.piconepress.com/projects/tuh_eeg/html/downloads.shtml}{TUH EEG Seizure Corpus 2.0.0} & TUSZ & 2022 & 675  & N/A \\ 
        \href{https://data.mendeley.com/datasets/5pc2j46cbc/1}{American university of Beirut Medical Center}& AUB-MC & 2021 & 6 & N/A  \\
        \rowcolor{Gray}    
        \href{https://physionet.org/content/siena-scalp-eeg/}{Siena Scalp EEG} & SSE & 2020 & 14  & 20 - 71 Y \\
        \href{https://zenodo.org/record/2547147\#.ZDkyPi_pNQI}{Helsinki University Hospital EEG} & HUH & 2018 & 79 & \makecell{35 - 45 w}  \\
        \rowcolor{Gray}    
    
        \href{https://www.researchgate.net/publication/308719109\_EEG\_Epilepsy\_Datasets}{Neurology and Sleep Centre Hauz Khas} & NSC-HK & 2016 & 10 & N/A \\
        \href{https://physionet.org/content/chbmit/1.0.0/}{CHB-MIT Scalp EEG Database}  & CHB-MIT & 2010 & 22 & 1.5 - 22 Y \\
             
        \bottomrule
    \end{NiceTabularX} 
    \caption{General information regarding public seizure scalp EEG datasets. In cases where published summaries lacked specific information, we examined the datasets to retrieve the information. N/A denotes that we failed to retrieve the information, even after manually inspecting the dataset.}
    \label{table1}
\end{table}

\begin{table}[H]
    \footnotesize
    \centering
    \begin{NiceTabularX}{\textwidth}{ccccccY}
        \toprule
        \textbf{Dataset} & \makecell{\textbf{Segment} \\ \textbf{duration}} & \makecell{\textbf{Number of} \\ \textbf{seizure events}} & \makecell{\textbf{Number} \\ \textbf{of segments}} & \makecell{\textbf{Sampling} \\ \textbf{frequency (Hz)}} & \makecell{\textbf{Continuous} \\ \textbf{data}} & \makecell{\textbf{Number of} \\ \textbf{channels}} \\
        
        \midrule
        \belowrulesepcolor{Gray}
        
        \rowcolor{Gray}
        TUSZ & \textbf{$^1$} & 4029 & 7377 & 250 \textbf{$^2$} & yes & 23 - 31\\

        AUB-MC & \textbf{$^1$} & 35 & 3895 & 500 & no & 19\\

        \rowcolor{Gray}
        SSE & \textbf{$^1$} & 47 & 41 & 512 & yes & 20 or 29 \\
         
        HUH & \textbf{$^1$} & 460 & 79 & 256 & yes & 19 \\
        
        \rowcolor{Gray}
        NSC-HK & 5.12s & \makecell{50} & 150  & 200  & no & 1 \\
        
        CHB-MIT & 1 - 4 h & 198 & 664 & 256 & yes & 18 - 26 \\

        %\aboverulesepcolor{Gray}
        \bottomrule
    \end{NiceTabularX}  
    \caption{Characteristics and recording parameters of the EEG segments of publicly available EEG datasets. Note that the NSC-HK dataset has one EEG channel and no information on how it was selected or which placement it corresponds to. In cases where published summaries lacked specific information, we examined the datasets to retrieve the information.\\ \textbf{$^1$} The recording lengths vary between segments.\\ \textbf{$^2$} This is the minimum sampling frequency. }
    \label{table3}
\end{table}

Important differences not presented in the tables include EEG electrode numbers, locations (see Section \ref{sEEG_section}), montage, EEG recording reference, and the initial pre-processing steps applied to the raw data. This heterogeneity poses significant challenges for making direct comparisons of algorithms trained on different datasets. 

Here we expand further on specific characteristics that vary across datasets, of which algorithm designers should be aware. 

\textbf{Seizure labelling and annotations}: A range of labelling systems have been used across the datasets to label segments of EEG. Following Section \ref{Seizure-related_phases}, non-seizure periods could correspond to pre-ictal, post-ictal or inter-ictal phases. However, only one dataset (HUH) distinguishes between pre-ictal, ictal and inter-ictal segments, with the rest using a binary seizure/non-seizure labelling system. For datasets with continuous data, it would be possible to extract pre-ictal and post-ictal activity by selecting segments surrounding the period labelled as seizure.

There are also differences in the granularity of seizure classes, ranging from a binary seizure/non-seizure label (in the CHB-MIT and SSE datasets) to the eight seizure subtypes listed above, used in the TUH dataset \cite{tiwary2022deep}. Notably, none of the datasets have annotated segments explicitly labelling EEG from \emph{status epilepticus}, even though the electrographic manifestations of \emph{status epilepticus} may differ to that of isolated seizures. 

Most often, labelling is applied to the entire recording, known as \emph{term-based} labelling, and not at an individual channel level, known as \emph{event-based} labelling. We refer to this distinction as the \emph{spatial labelling granularity}. The fine-grained event-based labelling is useful to chart the spatial and temporal propagation of seizure activity. It also has implications for the development of seizure detection algorithms \cite{shah2018temple}.

In addition to spatial granularity, seizure labels can be applied at different temporal resolutions. The TUH dataset provides precise demarcation of seizure onset/offset, other datasets label pre-defined intervals of EEG segments. For example, a 30 seconds segment containing a 10 second seizure would be labelled as ictal. Again, this has implications for algorithm design and performance evaluation. 

\textbf{Pre-processing}: There are discrepancies between the pre-processing steps applied to different datasets. For example, the Haus Khas and AUB-MC datasets are band-pass filtered between 0.5-70 Hz and 1/1.6-70 Hz respectively, a 50Hz notch filter was applied to the AUB-MC dataset, and the TUH dataset contains raw recordings. 

\textbf{Balancing and partitioning}: Most datasets contain imbalanced data, where ictal periods constitute a small proportion of the available data \cite{rahman2013addressing} (Table \ref{table3}). However, two datasets contain balanced data. The Hauz Khas dataset contains 50 pre-ictal, ictal and inter-ictal segments respectively. Similarly, the AUB-MC dataset contains 3,895 1 second ictal and non-ictal segments respectively. The AUB-MC and TUSZ datasets have pre-partitioned data into independent train and testing subsets, thereby simplifying the comparison of algorithm performance.

A range of techniques have been used to compensate for imbalanced datasets. Firstly, \citeauthor{jiang2023seizure} and \citeauthor{jiang2019transfer} use random undersampling of the majority non-seizure class to provide an equal number of seizure and non-seizure segments. We found this to be the most popular balancing strategy. Secondly, a weighted loss function can be used to alleviate the imbalance in the two classes and minimise biased learning \cite{khalkhali2021low, einizade2020deep, chatzichristos2020epileptic, thuwajit2021eegwavenet}. Thirdly, additional non-ictal segments can be created using generative adversarial networks (GAN) \cite{zhang2019epilepsy}. 

\newpage
\section{PRISMA: A systematic search strategy} \label{prisma_section}
\begin{figure}[H]
        \centering  \includegraphics[width=0.9\textwidth, trim={0 0 0 1cm},clip]{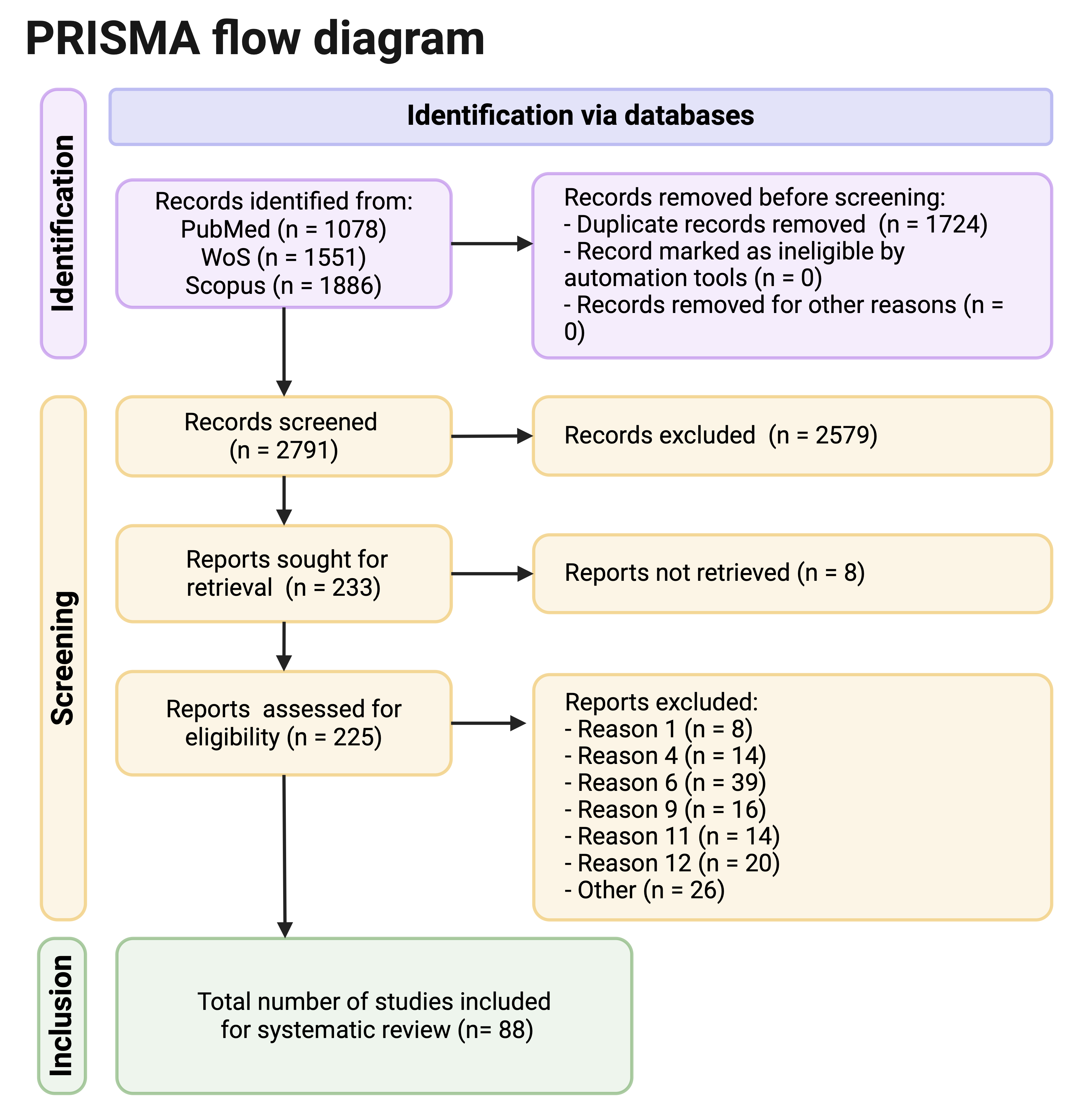}
        \caption{PRISMA flowchart. Reasons for exclusion of reports are numbered following the list of exclusion criteria provided above. `Other' includes retracted articles, studies that are not seizure detection and studies that did not meet reasonable quality standards. (Created with BioRender, \cite{created_with_biorender_prisma})}
        \label{prisma}
\end{figure}

We systematically selected 88 articles following the Preferred Reporting Items for Systematic Review and Meta-Analysis (PRISMA) guidelines \cite{page2021prisma}. The  strategy is summarised in Figure \ref{prisma}. The three databases searched were PubMed, Web of Science, and Scopus. The initial search was done on February 22, 2023 and the search strings for each database is shown in Table \ref{prisma_search_string} (Appendix). Due to the high number of relevant published material, the search was limited to content published from 2019 onward.

For each document identified, the title, abstract, keywords and author list were imported into Rayyan \cite{ouzzani2016rayyan}, software which facilitates the screening process. Out of 4515 initial entries, 1724 duplicates were removed, leaving 2791. Exclusion criteria included: (1) studies published before 2019, (2) studies that include other signals than EEG such as ECG or video data, (3) seizure prediction studies, (4) studies using only intracranial EEG data, (5) hardware implementation and testing studies, (6) patient specific algorithms, (7) clinical studies, (8) studies focusing on something else other than seizures, e.g., sleep staging, (9) studies exclusively using private data, (10) studies looking at a single type of epilepsy, (11) studies that do not include any ML, (12) studies that focus on data selected from specific patients without providing reasoning.

An initial screening to exclude articles was done using a keyword search and manual inspection of the results. Keywords screened for were: Bonn, prediction, clinical, sleep, video, neonatal, rats, rodent, survey, systematic review, review, mice, mouse, genetic, gene, Parkinson, MRI, stroke, IoT, antibodies, heart rate, patient specific, personalised, and COVID.

After this screening, 233 articles remained as potentially relevant and their full text manuscripts were retrieved. Out of these, 8 were not be retrieved as they were not free to access. Finally, a total of 88 records were included in the review. 

\newpage
\section{A review of automated seizure detection} \label{review_section}

\begin{figure}[h!]
        \centering        \includegraphics[width=\textwidth]{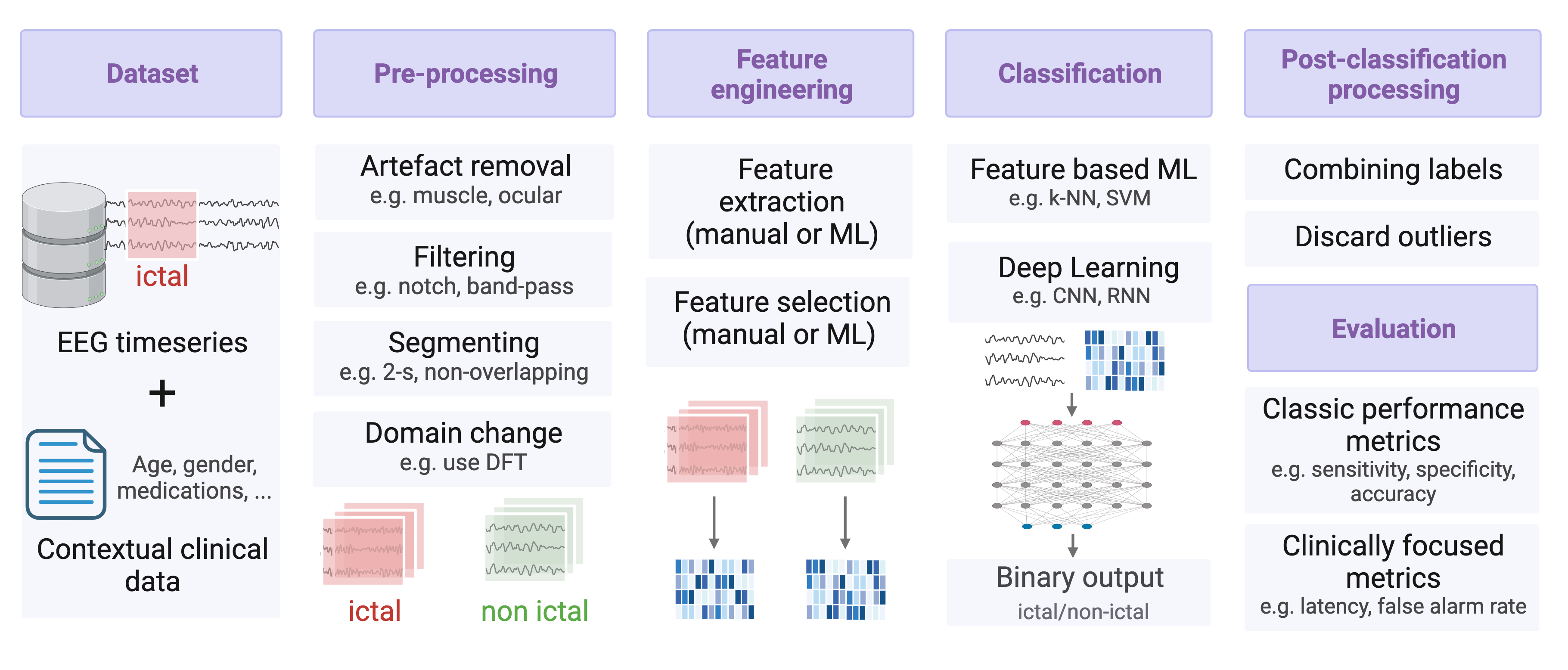}
        \caption{Standard pipeline for automated seizure detection using ML algorithms (Created with BioRender, \cite{created_with_biorender})}
        \label{fig:flowchart}
\end{figure}

Here we review the automated seizure detection ML literature. The stages of a traditional ML pipeline are summarised in Figure \ref{fig:flowchart}.  It usually begins with the application of data pre-processing steps, and ends with model evaluation.  In this section we comprehensively evaluate each possible stage of the pipeline. However, it is worth noting that not all algorithms necessarily adhere to every step of this pipeline.

\subsection{Data pre-processing}
Pre-processing encompasses a number of possible steps (see Figure \ref{fig:flowchart}). Scalp EEG data may be corrupted by a wide variety of noise sources and artefacts. Removing contaminated data or filtering out artefacts can therefore enhance the signal-to-noise ratio and data consistency, potentially improving seizure detection performance. However, pre-processing is a double-edged sword: there is a risk of removing informative signal, such as epileptiform activity. Furthermore, the run time of some pre-processing steps can reduce an algorithm's potential to operate in real-time. 

\subsubsection{Noise reduction and artefact removal}

A range of techniques have been applied to reduce the impact of noise, although none of the reviewed studies benchmark and compare the effectiveness of their techniques on classification performance.
    
Filtering techniques are applied to attenuate selected frequencies in the EEG. A filter can be designed to remove frequencies above a threshold (low-pass), below a threshold (high-pass), frequencies that are not within a chosen interval (band-pass), or to remove those that are (notch).

High-pass filtering is commonly employed to remove the effect of slow (< 1 Hz) presumed non-neural baseline signal drift. A low-pass filter is often used at the time of recording to prevent aliasing, which occurs when the sampling frequency is lower than twice the maximum frequency in the signal. Low-pass filtering is also often used offline to reject frequencies above ~30-45 Hz, considered to mainly contain irrelevant noise and artefacts. However, this may result in the exclusion of informative neurally-driven high-frequency signals \cite{gotman2010high, de2019filters}. In practice, band-pass filtering is generally applied to attenuate both low- and high-frequency components (e.g., 0.3-60 Hz \cite{humairani2022wavelet}). Notch filters are commonly used to remove power line interference (e.g. 60 Hz in North America and 50 Hz in Europe \cite{salafian2022cnn, salafian2021efficient, craley2019integrating, einizade2020deep, wei2020epileptic}). There is no consensus on the selection of frequencies to attenuate, with studies generally excluding frequencies outside the range of $\sim$ 0.5-40 Hz.

Beyond differences in bandwidths, a range of digital filtering techniques are employed. The Butterworth and the Savitzky–Golay filters are most commonly used. Less commonly encountered are adaptive filters, used for artefact removal. Their ability to continuously adapt to the input makes them well-suited for dealing with non-stationary data and hardware degradation. Notably, \citeauthor{poorani2021seizure} use an adaptive vector median filter to remove artefacts from the CHB-MIT data \cite{poorani2021seizure}.

A range of blind source separation (BSS) techniques have been applied to EEG to recover the original source signals from their observation (mixed with non-neural sources). This requires the different source signals to have different statistical properties. Commonly encountered BSS techniques are principal component analysis (PCA), independent component analysis (ICA), non-negative matrix factorisation (NNMF), and common spatial pattern (CSP). It is commonly applied to remove stereotypical artefacts. One example is the blinking artefact, which has non-Gaussian statistical properties independent of other EEG source signals, making it appropriate for removal using ICA \cite{jung2000removing}. For example,  \citeauthor{raghu2019performance} uses a notch filter and a 0.5-40 Hz band-pass filter before ICA-based artefact removal \cite{raghu2019performance}. However, these techniques can remove potentially useful EEG information.

Another technique used for pre-processing, but also for feature extraction, is the wavelet transform. A wavelet, unlike the sinusoids used in the Fourier transform, is a decaying, wave-like oscillation. This makes it better suited to non-stationary signal analysis \cite{kumar2008removal}. The wavelet transform evaluates how much of a wavelet is present in a signal for a specific scale and time shift, using convolution. It is sometimes used to separate the EEG into the canonical frequency bands (see definitions above) \cite{kumari2011seizure, bajaj2020wavelets}. Wavelet decomposition can help identify artefacts from each channel individually, as opposed to ICA/PCA approaches that require multi-channel data. Wavelet transforms can also be used as a set of new features. For example, \citeauthor{nemati2022medium} use the discrete wavelet transform (DWT) to decompose the EEG signal and subsequently produce a correlation map which is used as the input to a classifier \cite{nemati2022medium}.

% Remove capital letters for acronym expansions above me ^^

Breaking down signals into various components can be done using BSS techniques, Fourier transform, and Wavelet transforms. It can also be performed using empirical mode decomposition (EMD) and variational mode decomposition (VMD). These methods decompose a signal into various time domain signals called intrinsic mode functions (IMFs) \cite{molla2012artifact}. Whilst EMD decomposes the signal adaptively in a data-driven way, VMD uses a non-recursive decomposition technique. Even though both methods can handle non-linear data, EMD is the only one capable of working on non-stationary signals \cite{ju2020variational}. For these reasons, EMDs are widely used in biological signal analysis  \cite{bajaj2011classification, singh2017fourier}. Raw IMFs or features extracted from IMFs can be used as input to a ML algorithm. For example, \citeauthor{ru2022epilepsy} employ VMD to partition the signal originating from electrodes FZ-CZ and CZ-PZ into five distinct IMFs. These IMFs' phase synchrony index is then used as input for a detection model \cite{ru2022epilepsy}. In a related study, \citeauthor{davidson2022seizure} compare three distinct source isolation techniques for artefact removal: EMD, empirical wavelet transform (EWT), and variational mode decomposition (VMD) \cite{davidson2022seizure}. The EMD method showed the best performance. 

Numerous EEG pre-processing approaches exist, which can all be combined in multiple ways. There is no standard method for pre-processing EEG data. We recommend that researchers  apply their detection algorithm on raw data, data that has undergone minimal filtering, and data that has been pre-processed with more complex techniques like ICA or wavelet decomposition. This way, performance improvements due to different pre-processing steps can be clearly measured in the context of seizure detection. For example, it is frequently assumed that DL techniques handle raw data better than feature-based ML, however, we found no study that tested this assumption quantitatively in the context of seizure detection.

\subsubsection{Channel selection}

The electrical activity of the brain is recorded by scalp EEG, typically using around 20 electrodes. Selecting a subset of these electrodes for analysis and disregarding others can decrease an algorithm's training time due to a reduction in the input dimension. However, there is a risk of dismissing clinically relevant information, hereby decreasing the algorithm's performance and generalisability. Here we outline the different approaches to channel selection. 

Channel selection can be performed based on someone's critical judgment and assessment of which electrodes are, in theory, most useful to keep. For instance, \citeauthor{wei2020epileptic} select 4 channels of the TUSZ data capable of recording different brain regions: F8-T4 (temporal lobe),  T5-O1 (occipital lobe), FP1-F3 (frontal lobe) and F3-C3 (central cerebral hemisphere) \cite{wei2020epileptic}. On the other hand, \citeauthor{ru2022epilepsy} only use channels FZ-CZ and CZ-PZ \cite{ru2022epilepsy}. The examples above illustrate theory-based channel selection. 

Alternatively, some studies rely on quantitative approaches to channel selection in order to minimise the loss of information and drop in performance. For instance,  \citeauthor{nandini2022efficient} use PCA to perform channel selection \cite{nandini2022efficient}. Although useful to reduce algorithmic complexity, this data-centric selection process is often not applicable in real-time, which limits its relevance to clinical applications. Additionally, seizure activity may propagate to different brain regions, which might be missed if the channel selection process is performed using a short segment of EEG. 

It is also important to identify and state the motivation behind channel selection, whether it is to decrease the run time, to reduce redundancy, or to make the clinical implementation more realistic. The purpose of channel selection should be the identification of a subset of channels that are patient independent and contain enough information to identify all types of seizures. We also recommend the channel selection process to be repeated across different datasets, to ensure that the identified channels are dataset independent.

\subsubsection{Data segmentation}

Segmenting data is important for the development of online algorithms that are capable of making real-time decisions. There are two parameters involved in the segmentation process: the segment length and amount of overlap between consecutive segments. We identify three important clinical implications for choosing these parameters.

\begin{itemize}
    \item  Given the non-stationarity of seizure activity, the length of the segment used influences the patterns the model learns. Most algorithms use EEG segments shorter than 10 seconds as input, as they are generally short enough to be considered quasi-stationary. 

    \item The amount of overlap between the segments can be leveraged to (1) reduce the imbalance in the training data, and (2) reduce detection latency during inference. Introducing overlap between the extracted segments increases the number of samples. This can help to train the model, but can also introduce redundancy, and lead to overfitting. A small number of studies have used non-overlapping EEG segments lasting at least 60 seconds for training \cite{sharma2022fractal, kaziha2020convolutional, chen2020epilepsy}. During inference, the longer the time between the analysis of two consecutive EEG segments, the larger the seizure detection alarm delay can be. Note that using non-overlapping segments during training does not prevent the use of overlapping segments during inference, which can help in reducing potential alarm delays. The length of the EEG segments used in encountered studies and whether they are overlapping is shown in Table \ref{ML_1} to \ref{DL5}. 

    \item A segment length which is too short can remove the historical context required for an accurate classification. For context, clinicians usually read EEG signals in 10–15 seconds long pages, sometimes requiring viewing multiple pages for accurate detection. This context can be incorporated by some ML algorithms with memory such as recurrent neural networks (RNNs), or, by using some post-classification analysis (Section \ref{Post-classification processing}). For instance, the detection of status epilepticus depends on having 5 minutes of ictal activity, which a single 5 second segment does not provide.

\end{itemize}

\subsubsection{Validation strategy}\label{validation_section}

Model training and evaluation requires the data to be divided into either a train-set and a test-set or into a train-set, validation-set and test-set. Each of these sets serves a different purpose in the model building pipeline. The train-set is the data used during model training. The validation-set is used after each training epoch to evaluate the performance of the model. It further enables tuning of the model's hyperparameters without overfitting to the train data. The test-set is used to evaluate the performance of the model on unseen data after training.

There are a number of ways to construct either the train-test or the train-validation-test datasets. Common examples are  (1) random split, (2) k-fold cross-validation and (3) leave-one-out validation. All of these approaches can be used for train-test or train-validation-test splits. Tables \ref{ML_1}-\ref{ML_3} and \ref{DL}-\ref{DL5} summarise the validation strategy of the reviewed literature.

A random split is the simplest data splitting strategy. In this regime, the data is randomly divided into 2 or 3 different sets, often with 10\% for testing, and the remaining 90 \% solely for training, or for training and validation.

K-fold cross-validation (CV) partitions the dataset in k evenly-sized pieces, each of which is called a fold. K is often chosen as 5 or 10. On each iteration, one of the k pieces is taken as the test set and the remaining data, that is, the other k-1 pieces, are taken as the training set, or the training and validation sets. A model is then trained and tested  using these datasets and the performance recorded. The performances of the k models are then averaged, which gives a more reliable estimate of the model's performance on unseen data.

Finally, the leave-one-out (LOO) strategy, also called leave-one-patient-out in the context of seizure detection, is the same in principle as k-fold CV, but where one takes the data of a single patient as the test set, and the remaining data as the training or training and validation sets. It is a special case of k-fold CV.  We refer to this as LOO cross validation (LOO-CV). As an example, the CHB-MIT dataset contains 22 subjects, so using LOO-CV would imply training 22 models, each using the data of 21 patients and testing on the data of the patient that was `left out'. Each model is tested using the data of a different patient. Patient independent performance of the model is obtained by averaging the performance of the 22 models. This is a rigorous validation strategy in the field of seizure detection as it prevents overfitting and is patient independent. When using datasets containing a large numbers of subjects, or when the amount of data per patient is scarce, the leave-one-group-out CV (LOGO-CV) strategy may be more suitable. This is similar to LOO-CV but uses the data of a group of patient as the test set, as opposed to using the data of a single patient.

\subsection{Feature engineering} 

A feature of a time-series is any calculated measure, for example, the power in set frequency band, entropy values or Hjorth parameters. This section starts by describing the different EEG domains that features can be extracted from, followed by how this set of features can be reduced to an adequate subset for classification purposes.

\subsubsection{Data domains} 

The domains, or representations, which EEG time-series data can be mapped to are: time, frequency, time-frequency, and network. In this subsection, we briefly review the four different domains for EEG data representation.

The time domain is simply the amplitude as a function of time, and is the domain of raw EEG recordings. 

The frequency domain refers to the representation of signals in terms of their frequency content. Switching from the time to frequency domain can be achieved using the Fourier transform or Laplace transform (for the complex frequency domain).

The time-frequency domain represents how the frequency components of a signal change over time. This is evaluated by segmenting the data into consecutive segments whose frequency components are extracted one after the other. The output is 2-dimensional and known as a spectrogram. Various techniques exist to obtain a time-frequency representation of a signal, but common ones include the short-time Fourier transform (STFT) and continuous wavelet transform (CWT). Spectrograms are often used as input to convolutional neural networks (CNNs) and are widely used in seizure detection. 

Finally, EEG data can be represented in a network domain, capturing \textit{functional connectivity}, or measures of how brain regions interact with one another over time. Typically, a network is constructed as a weighted connected graph, where each EEG channel is represented by a node and edges between nodes are assigned numerical quantities corresponding to the connectivity between the signals from pairs of EEG channels. For instance, \citeauthor{li2022graph} used the cross-correlation between channels as edge values \cite{li2022graph}, but a range of connectivity-based measures can be used, including phase synchrony and the phase lag index. Once the EEG data is represented as a connected graph, a range of graph theoretical metrics, such as the degree and clustering coefficient, can be used as features.

\subsubsection{Features} 
Table \ref{feature_table} shows commonly extracted features from EEG for seizure detection. Importantly, EEG features are either univariate, bivariate, or multivariate, that is, involving one, two, or more than two channels in the analysis, respectively \cite{andrzejak2023seizure}.

Although features are typically applied to EEG data in a certain domain, as shown in the table, they can be applied to other domains, provided that the dimension of the input is compatible with the arithmetic steps of the feature calculation. For example, Hjorth parameters are defined for the time domain. However, it is possible to calculate them for any series data, such as a power spectrum which belongs to the frequency domain. Of course, the interpretation of features differs according to the domain of the data to which they are applied.

\begin{table}[H]
    \centering
    \scriptsize
    \begin{tabularx}{\linewidth}{p{0.25\textwidth}<{\centering}|p{0.7\textwidth}<{\centering}}
            \toprule       
            \textbf{Domain} & \textbf{Features} \\
            
            \hline
            \multirow{5}{*}{\textbf{Time domain}} & \cellcolor{Gray}Absolute energy, energy ratio \\
            
            &  Mean and variance over the autocorrelation for different lags, partial autocorrelation \\
            
            & \cellcolor{Gray} Autoregressive coefficients \\
           
            & Hjorth activity, mobility, and complexity, line length  \\
            
            & \cellcolor{Gray} Mean, median, mode, maximum, minimum, mean, standard deviation, variance, quantile range, kurtosis \\

            & Permutation entropy, Shannon entropy, spectral entropy, approximate entropy, Sample entropy, Fuzzi entropy, Renyi entropy\\
            
            \hline \textbf{Frequency domain} &  \cellcolor{Gray} Fourier transform aggregate and coefficients, band power \\
            
            \hline \textbf{Time-frequency domain} & Continuous wavelet coefficients and peaks \\
    
            \hline \textbf{Network domain} & \cellcolor{Gray} Weighted degree, clustering coefficient, harmonic centrality, modularity, closeness centrality, eigenvector centrality \\
            
            \bottomrule
    \end{tabularx}
    \caption{Example of popular time-series features used to discriminate between ictal and non-ictal EEG segments. Even though features can be extracted from data belonging to any domain, providing it has the correct dimensions, they are often used in one domain only. The features in the table below are grouped by the domain within which they are most commonly applied.} 
    \label{feature_table}
\end{table}

\subsubsection{Feature selection}

It is common to encounter studies where a large number of features from different domains are extracted. Only using a subset of those features as input to a ML algorithm can reduce computational complexity, redundancy, and bias. 

Broadly, there are four strategies for feature selection. Firstly, dimensionality reduction is commonly achieved through principal component analysis (PCA). PCA can be applied prior to training the algorithm to reduce the dimensionality of either the EEG input signal directly or the input feature space, as done by \citeauthor{jiang2023seizure} \cite{jiang2023seizure}. Other algorithms reducing the dimensionality of the feature space include minimal redundancy maximal relevance (mRMR). mRMR emphasizes selecting a feature subset that optimally balances relevance to the target variable while minimizing redundancy among the selected features.  

Secondly, after an algorithm has been trained, feature selection can be informed by an evaluation of feature importance, that is, the relative weights of each feature in a classification process. For example, \citeauthor{wu2020detecting} extracted a total of 798 features from the raw EEG signal and several IMFs obtained after complete ensemble empirical mode decomposition (CEEMD) \cite{wu2020detecting}. Using an XGBoost algorithm allowed them to quantify which of the 798 features were allocated the highest discriminative power, or rank. Among the 20 features with the highest rank obtained when using the CHB-MIT dataset, 14 belong to the time domain (70\%), 4 belong to the frequency domain (20\%), 1 belongs to the time-frequency domain (5\%) and 1 is an entropy-based feature (5\%). A similar analysis was performed on the Bonn dataset and revealed that although some features were similarly high ranking across both datasets, others were not. For instance, the autoregressive coefficients of the raw signal and IMF10 trace both played an important role in the classification of the Bonn dataset. However, they were not among the 20 most important features when analysing the CHB-MIT dataset. This analysis shows that features from multiple domains can contribute to differentiating ictal and non-ictal EEG segments. It also demonstrates the heterogeneity of publicly available datasets, and the challenges of developing a patient and hardware independent algorithm. Other methods for quantifying feature importance are Shapley's additive explanations (SHAP) \cite{marcilio2020explanations}, and sequential forward feature selection (SFFS). In SFFS, features are iteratively added to the model and the corresponding improvement of performance metrics like accuracy or precision is evaluated at each step.

Thirdly, features can be chosen according to the required explainability of the algorithm. The correspondence of some features to ictal activity is intuitive: for example, it is easy to understand that, generally, ictal activity is associated with an increase in the power of the EEG signal. However, many features discriminate in less intuitive ways, e.g. auto-regressive coefficients. Using features where the meaning is difficult to understand, or using too many features, can make the results difficult to interpret or risk overfitting the data. This could generate scepticism amongst end-users and pose a barrier to medical deployment.

Fourthly, choosing which features to use can be based upon their computational complexity and, hence, the time they require to be calculated. For example, the complexity of approximate entropy and sample entropy are both \(\mathcal{O}(n^2)\), where \(n\) is the length of the time series. As such, they are less suitable for real-time  applications using EEG segments with a large number of data points \cite{qiu2022lightseizurenet}. On the other hand, the fast Fourier transform has a computational complexity of \(\mathcal{O}(n\log(n))\). The difference in compute time required to calculate a feature that scales linearly with respect to sample size as opposed to one that scales quadratically with respect to sample size becomes more significant as \(n\) increases. Therefore, features associated with a lower computational complexity, such as the power present in the canonical bands, may be better suited to online seizure detection than features associated with a higher computational complexity, especially for large \(n\).

\subsection{Feature-based ML algorithms for classification}
Once the data pre-processing, feature extraction, and feature selection steps are completed, the data is ready for classification. It is worth noting that deep learning (DL) does not necessarily require extracted features as input. This section reviews three common feature-based ML algorithms for automated seizure detection. We focus on  DL models in the following section. Tables \ref{ML_1}, \ref{ML_2}, \ref{ML_3} summarise the feature-based ML models encountered in the literature.

Firstly, k-nearest neighbours (k-NN) is one of the simplest supervised machine learning algorithms used in seizure detection. For classification tasks, the algorithm assigns the label to the test data that is most common amongst its k-nearest neighbours. As this algorithm necessitates data points associated with labels as input, it does not support time series data. Therefore, features of the EEG segments are used as input for the classification of ictal and non-ictal activity. For instance, \citeauthor{jumaah2020epileptic} perform a discrete cosine transform on the EEG segments. The energies of the sub-bands of the power spectrum are the input to a k-NN algorithm which achieved an accuracy of 93.64\% using 5 of the 23 available channels of the CHB-MIT dataset \cite{jumaah2020epileptic}. \citeauthor{jiang2019transfer} extract features using a dual-tree discrete wavelet transform and use a feature selection method known as transfer component analysis \cite{jiang2019transfer}. The binary k-NN had an overall accuracy of 74.03\% on the CHB-MIT dataset. A downside of the k-NN algorithm is that it is a non-parametric method, meaning that it does not fit a model during training. Instead, it `memorises' the training data.  Identifying the k nearest neighbours of a new data point requires computing its distance to every training data points available. This computational inefficiency makes k-NN too slow for online applications using large amounts of data \cite{marimont1979nearest, berisha2021digital}. 

A second popular algorithm which can handle larger datasets is the support vector machines (SVM). It is a parametric method that finds the optimal hyperplane separating the data into various groupings. \citeauthor{humairani2022wavelet} compares the use of Shannon entropy and Renyi entropy as input to a SVM \cite{humairani2022wavelet}. They achieve a 92.96\% accuracy using Renyi's entropy on the CHB-MIT dataset. Similarly, \citeauthor{raghu2019performance} use a DWT based sigmoid entropy as input to a SVM to perform binary classification on the CHB-MIT data and achieve a 94.21\% accuracy \cite{raghu2019performance}. \citeauthor{fathima2020wavelet} use more features as input to an SVM, they first compute the DWT decomposition at level 5 using Daubechies 4 wavelet. Note that the chosen wavelet should be the one whose profile best matches the characteristic shape we are looking to detect in the EEG. The Daubechies wavelets are well known for their resemblance to biological signals, and are often used in DWT decomposition of EEG. The level, or scale, of decomposition refers to how `stretched' the wavelet is. The higher the level, the wider the wavelet. Choosing the scale of decomposition is also arbitrary and depends on the nature of the signal. Back to \citeauthor{fathima2020wavelet}, who calculate the mean, RMS, inter-quartile range, and entropy of the decomposed signal  \cite{fathima2020wavelet}. The reported accuracy is 98.6\%, however, there is no justification behind the restriction of the feature extraction to the level 5 decomposition of the EEG signal. Some papers use algorithms for feature extraction and selection prior to using them as input to the SVM for classification. This is the case of \citeauthor{shariat2021automatic}, who compare two feature selection techniques; sequential forward feature selection (SFFS) and minimal redundancy maximal relevance (mRMR) \cite{shariat2021automatic}. They achieve a maximal accuracy of 99.87\% using SFFS followed by a SVM on all 23 patients of the CHB-MIT.  

The third common feature-based ML algorithm for seizure detection is the gradient boosting machine (GBM). This is an ensemble learning method that consists of multiple weak prediction models, typically decision trees, that are combined to form a stronger prediction model. An initial weak sub-learner is trained and, subsequently, a second sub-learner is constructed to fit the residuals of the first one, et cetera. \citeauthor{nandini2022efficient} calculate seven time domain features (min/max, median, mean, skewness, standard deviation and kurtosis) of EEG signals decomposed using wavelet transforms. For classification, they use an extreme gradient boosting (XGBoost) algorithm, and achieve a classification accuracy of 94.46\% \cite{nandini2022efficient}. 

In addition to the three main classification algorithms, other less common methods include random forest classifiers (RF) and linear discriminant analysis (LDA) \cite{xiong2023seizure, rohira2023automatic, ansari2021patient, slimen2020eeg, he2021progressive, ru2022epilepsy}. For instance, \citeauthor{xiong2023seizure} use a RF classifier on the CHB-MIT and Siena datasets using network features such as weighted degree and clustering coefficient. Their reported accuracy is 84.83\% \cite{xiong2023seizure}. \citeauthor{rohira2023automatic} compare two inputs to a RF classifier, (i) spectral-based measure of functional connectivity, and (ii) the power of 6 frequency bands \cite{rohira2023automatic}. Using 8 channels they achieve an accuracy of 90.87\% and 95.73\% when using coherence coefficients and the power of different frequency bands respectively.  

% ML tables
\newpage
\begin{table}[H]
    \footnotesize	
    \centering
    \begin{adjustbox}{width=1\textwidth}
    %\scriptsize
        \begin{tabular}{p{0.2\textwidth}<{\centering}p{0.3\textwidth}<{\centering}ccp{0.15\textwidth}<{\centering}ccp{0.1\textwidth}<{\centering}}
        \toprule 
        \textbf{Classifier} & \textbf{Feature(s)} & \textbf{Dataset(s)} & \textbf{Performance}  & \textbf{Validation} & \textbf{\makecell{Segment\\Length}}& \textbf{Year} & \textbf{Reference} \\
        \midrule
        \belowrulesepcolor{Gray}
        
        \rowcolor{Gray}
        
        \adjustbox{stack=cc, margin=0ex 2ex}{k-NN}& \adjustbox{stack=cc, margin=0ex 2ex}{Energy of signal after DCT} & \adjustbox{stack=cc, margin=0ex 2ex}{CHB-MIT\\(21 patients,\\5 electrodes)}& \adjustbox{stack=cc, margin=0ex 2ex}{acc: 93.64\%\\sen: 94.77\%\\spe: 92.21\%\\F-score: 93.12\%\\FPR: 0.07\\FNR: 0.05\\Error: 0.06}  & \adjustbox{stack=cc, margin=0ex 2ex}{10-fold\\CV} & \adjustbox{stack=cc, margin=0ex 2ex}{1s,\\no overlap} & 2019  & \adjustbox{stack=cc, margin=0ex 2ex}{\cite{jumaah2020epileptic}}  \\

        \adjustbox{stack=cc, margin=0ex 2ex}{fuzzy k-NN} & \adjustbox{stack=cc, margin=0ex 2ex}{GNMF decomposed SSTFT maps} & \adjustbox{stack=cc, margin=0ex 2ex}{CHB-MIT,\\Bonn}& \adjustbox{stack=cc, margin=0ex 2ex}{acc: 98.99\%,\\sen: 99.27 \%,\\spe: 98.53\%}  & \adjustbox{stack=cc, margin=0ex 2ex}{10-fold\\CV} & \adjustbox{stack=cc, margin=0ex 2ex}{1s,\\no overlap} & 2023 & \adjustbox{stack=cc, margin=0ex 2ex}{\cite{li2023gnmf}}  \\

        \rowcolor{Gray}

        \adjustbox{stack=cc, margin=0ex 2ex}{Neutrosophic logic k-NN} & \adjustbox{stack=cc, margin=0ex 2ex}{\(\theta,\beta,\delta,\alpha\) power bands of four wavelet bands and \(\alpha\) to \(\delta\) power band ratio} & \adjustbox{stack=cc, margin=0ex 2ex}{Bonn,\\CHB-MIT (18 channels,\\no patient 12),\\AIIMS} & \adjustbox{stack=cc, margin=0ex 2ex}{acc: 89.06\%,\\ sen: 85\%,\\spe: 89.06\%}  & \adjustbox{stack=cc, margin=0ex 2ex}{LOO} & \adjustbox{stack=cc, margin=0ex 2ex}{1s,\\no overlap} & 2020 & \adjustbox{stack=cc, margin=0ex 2ex}{\cite{ansari2020automatic}}\\ 

        \adjustbox{stack=cc, margin=0ex 2ex}{k-NN\\ (feature selection),\\ RF\\(classification)} & \adjustbox{stack=cc, margin=0ex 2ex}{
        Weighted degree, clustering coefficient} & \adjustbox{stack=cc, margin=0ex 2ex}{CHB-MIT,\\Siena scalp}& \adjustbox{stack=cc, margin=0ex 2ex}{CHB-MIT:\\ F1: 86.69\% \\ AUC: 84.33\% \\ acc: 84.83\% \\ pre:85.60\% \\ sen :87.81\% \\ spe: 81.01\%}  & \adjustbox{stack=cc, margin=0ex 2ex}{5-fold\\CV} & \adjustbox{stack=cc, margin=0ex 2ex}{4s} & 2023 & \adjustbox{stack=cc, margin=0ex 2ex}{ \cite{xiong2023seizure}}  \\

        \rowcolor{Gray}

        \adjustbox{stack=cc, margin=0ex 2ex}{TCA (feature mapping\\ to latent sub-space)\\and k-NN\\(classification)} & \adjustbox{stack=cc, margin=0ex 2ex}{Separable features from dual-tree discrete wavelet parameters} & \adjustbox{stack=cc, margin=0ex 2ex}{CHB-MIT}& \adjustbox{stack=cc, margin=0ex 2ex}{acc: 74.03\%\\F1: 0.7473\\AUC: 0.8204}  & \adjustbox{stack=cc, margin=0ex 2ex}{LOO} & \adjustbox{stack=cc, margin=0ex 2ex}{3s,\\2.5s overlap}  & 2019 & \adjustbox{stack=cc, margin=0ex 2ex}{\cite{jiang2019transfer}}  \\

        \adjustbox{stack=cc, margin=0ex 2ex}{SVM vs k-NN} & \adjustbox{stack=cc, margin=0ex 2ex}{Hurst exponent, Logarithmic HFD}  & \adjustbox{stack=cc, margin=0ex 2ex}{CHB-MIT}& \adjustbox{stack=cc, margin=0ex 2ex}{SVM\\acc: 99.81\%,\\rec: 100\%,\\TNR: 0.99\\ k-NN\\acc: 93.21\%,\\rec: 92.56\%,\\TNR: 0.92}  & \adjustbox{stack=cc, margin=0ex 2ex}{10-fold\\CV} & \adjustbox{stack=cc, margin=0ex 2ex}{300s,\\240s overlap} & 2022 & \adjustbox{stack=cc, margin=0ex 2ex}{\cite{sharma2022fractal}}  \\

        \rowcolor{Gray}

        \adjustbox{stack=cc, margin=0ex 2ex}{SVM,\\LDA,\\k-NN}& \adjustbox{stack=cc, margin=0ex 2ex}{Power, mean, std extracted after using DTCWT}  & \adjustbox{stack=cc, margin=0ex 2ex}{CHB-MIT}& \adjustbox{stack=cc, margin=0ex 2ex}{acc: 100\%}  & \adjustbox{stack=cc, margin=0ex 2ex}{10-fold\\CV} & \adjustbox{stack=cc, margin=0ex 2ex}{8s,\\overlap} & 2020 & \adjustbox{stack=cc, margin=0ex 2ex}{\cite{slimen2020eeg}}  \\

        \adjustbox{stack=cc, margin=0ex 2ex}{SVM, ELM\\(SVM is best)}& \adjustbox{stack=cc, margin=0ex 2ex}{Weighted FPE complexity-based feature (W-FPE-F)} & \adjustbox{stack=cc, margin=0ex 2ex}{CHB-MIT\\(12 patients),\\ Bonn}& \adjustbox{stack=cc, margin=0ex 2ex}{acc: 98.9883\%\\ spe: 89.3300\%\\sen: 94.1650\%}  & \adjustbox{stack=cc, margin=0ex 2ex}{10-fold\\CV} & \adjustbox{stack=cc, margin=0ex 2ex}{4s,\\3s overlap} & 2019 & \adjustbox{stack=cc, margin=0ex 2ex}{\cite{zhang2019novel}}  \\

        \rowcolor{Gray}

        \adjustbox{stack=cc, margin=0ex 2ex}{SVM}& \adjustbox{stack=cc, margin=0ex 2ex}{Shannon Entropy Renyi's Entropy after DWT} & \adjustbox{stack=cc, margin=0ex 2ex}{CHB-MIT}& \adjustbox{stack=cc, margin=0ex 2ex}{acc: 92.96\%\\(using Renyi)}  & \adjustbox{stack=cc, margin=0ex 2ex}{N/A} & \adjustbox{stack=cc, margin=0ex 2ex}{10 mins} & 2022 & \adjustbox{stack=cc, margin=0ex 2ex}{\cite{humairani2022wavelet}}\\

        \adjustbox{stack=cc, margin=0ex 2ex}{SVM}& \adjustbox{stack=cc, margin=0ex 2ex}{Std, mean absolute deviation, RMS, min, interquartile range, skewness, entropy and max were extracted over wavelet coefficients}  & \adjustbox{stack=cc, margin=0ex 2ex}{CHB-MIT}& \adjustbox{stack=cc, margin=0ex 2ex}{spe: 100\%\\ sen: 97.2\%\\  acc: 98.6\%\\}  & \adjustbox{stack=cc, margin=0ex 2ex}{N/A} & \adjustbox{stack=cc, margin=0ex 2ex}{5s} & 2020 & \adjustbox{stack=cc, margin=0ex 2ex}{\cite{fathima2020wavelet}}\\
        
        \rowcolor{Gray}

        \adjustbox{stack=cc, margin=0ex 2ex}{SVM} & \adjustbox{stack=cc, margin=0ex 2ex}{DWT based sigmoid entropy (in time and frequency domain)} & \adjustbox{stack=cc, margin=0ex 2ex}{CHB-MIT,\\Bonn,\\RMCH}& \adjustbox{stack=cc, margin=0ex 2ex}{sen: 94.21\%}  & \adjustbox{stack=cc, margin=0ex 2ex}{LOO} & \adjustbox{stack=cc, margin=0ex 2ex}{1s}  & 2019 & \adjustbox{stack=cc, margin=0ex 2ex}{\cite{raghu2019performance}}  \\

        \adjustbox{stack=cc, margin=0ex 2ex}{SVM} & \adjustbox{stack=cc, margin=0ex 2ex}{Successive decomposition index (SDI)} & \adjustbox{stack=cc, margin=0ex 2ex}{RMCH,\\CHB-MIT,\\TUSZ}& \adjustbox{stack=cc, margin=0ex 2ex}{CHB-MIT:\\sen: 97.28\%\\FA: 0.57/h\\median latency: 1.7s\\TUSZ:\\sen: 95.80\%\\FA: 0.49/h\\median latency: 1.5s}  & \adjustbox{stack=cc, margin=0ex 2ex}{LOO} & \adjustbox{stack=cc, margin=0ex 2ex}{1s,\\no overlap}  & 2019 & \adjustbox{stack=cc, margin=0ex 2ex}{\cite{raghu2020automated}}  \\
        
        \rowcolor{Gray}

        \adjustbox{stack=cc, margin=0ex 2ex}{SVM} & \adjustbox{stack=cc, margin=0ex 2ex}{Kurtosis, skewness, line length, quartile values, correlation coefficient matrix of the frequency energy between any two channels. PCA reduction of the dimensionality.} & \adjustbox{stack=cc, margin=0ex 2ex}{CHB-MIT,\\ Siena} & \adjustbox{stack=cc, margin=0ex 2ex}{acc: 96.67\%,\\spe: 95.62\%,\\sen: 97.72\%}  & \adjustbox{stack=cc, margin=0ex 2ex}{Bootstrap} & \adjustbox{stack=cc, margin=0ex 2ex}{1s, 0.5s\\overlap} & 2023 & \adjustbox{stack=cc, margin=0ex 2ex}{\cite{jiang2023seizure}}\\ 
        
        \aboverulesepcolor{Gray}
        \bottomrule
              
    \end{tabular}
    \end{adjustbox}
    \caption{Feature-based ML methods from systematic review of literature for seizure detection in scalp EEG data}
    \label{ML_1}
\end{table} 

\newpage

\begin{table}[H]
    \footnotesize	
    \centering
    \begin{adjustbox}{width=1\textwidth}
    %\footnotesize
    \begin{tabular}{p{0.2\textwidth}<{\centering}p{0.3\textwidth}<{\centering}ccp{0.15\textwidth}<{\centering}ccp{0.1\textwidth}<{\centering}}
        \toprule 
        \textbf{Classifier} & \textbf{Feature(s)} & \textbf{Dataset(s)} & \textbf{Performance}  & \textbf{Validation} & \textbf{\makecell{Segment\\Length}} & \textbf{Year} & \textbf{Reference} \\
        \midrule
        %\belowrulesepcolor{Gray}
        %\rowcolor{Gray}

        \adjustbox{stack=cc, margin=0ex 2ex}{SVM (classification), Sequential Forward Feature Selection (SFFS) vs Minimal Redundancy Maximal Relevance (mRMR) (feature selection)} & \adjustbox{stack=cc, margin=0ex 2ex}{Covariance matrices of channels modified using Riemannian geometry} & \adjustbox{stack=cc, margin=0ex 2ex}{CHB-MIT\\(22 channels)}& \adjustbox{stack=cc, margin=0ex 2ex}{CHB-MIT using SFFS\\and 10-fold CV:\\acc: 99.87\%\\sen: 99.91\%\\spe: 99.82\%}  & \adjustbox{stack=cc, margin=0ex 2ex}{10-fold\\CV} & \adjustbox{stack=cc, margin=0ex 2ex}{2s,\\no overlap} & 2021 & \adjustbox{stack=cc, margin=0ex 2ex}{\cite{shariat2021automatic}}  \\

        \rowcolor{Gray}
        
        \adjustbox{stack=cc, margin=0ex 2ex}{LS-SVM} & \adjustbox{stack=cc, margin=0ex 2ex}{Mean, std, variance, Shannon entropy, approximate entropy, spectral centroid, spectral speed, spectral flatness, spectral slope, spectral entropy, Hurst exponent, Katz fractal exponent} & \adjustbox{stack=cc, margin=0ex 2ex}{CHB-MIT} & \adjustbox{stack=cc, margin=0ex 2ex}{acc: 98.37\%,\\sen: 91.11\%,\\pre: 91.67\%, \\spe: 91.46\%,\\ADR: 91.28\%,\\G-mean: 91.28\%,\\AUC: 0.992}  & \adjustbox{stack=cc, margin=0ex 2ex}{10-fold\\CV} & \adjustbox{stack=cc, margin=0ex 2ex}{N/A} & 2022  & \adjustbox{stack=cc, margin=0ex 2ex}{\cite{mohapatra2022esa}}\\ 
        
        \adjustbox{stack=cc, margin=0ex 2ex}{SVM, Quadratic Discriminant\\Analysis (QDA) and LDA} & \adjustbox{stack=cc, margin=0ex 2ex}{Maximum value, minimum value, mean value, variance, range, skewness, kurtosis, estimation of cross-correlation and various connectivity measures from graph theory} & \adjustbox{stack=cc, margin=0ex 2ex}{CHB-MIT\\(14 channels)} & \adjustbox{stack=cc, margin=0ex 2ex}{accuracies:\\SVM: 98.09\%,\\QDA: 81.49\%,\\LDA: 80.90\%,\\SVM AUC: 99.7\%,\\SVM sen: 98.1\%\\SVM spe: 98.1\%}  & \adjustbox{stack=cc, margin=0ex 2ex}{10-fold\\CV} & \adjustbox{stack=cc, margin=0ex 2ex}{5s,\\no overlap} & 2022 & \adjustbox{stack=cc, margin=0ex 2ex}{\cite{ashokkumar2022application}}\\ 

        \rowcolor{Gray}

        \adjustbox{stack=cc, margin=0ex 2ex}{Layered directed acyclic graph support vector machine (LDAG-SVM)} & \adjustbox{stack=cc, margin=0ex 2ex}{Entropy, largest Lyapunov exponent, correlation dimension} & \adjustbox{stack=cc, margin=0ex 2ex}{CHB-MIT,\\Bonn}& \adjustbox{stack=cc, margin=0ex 2ex}{acc: 95\%\\ sen: 99\%\\spe: 96\%\\run time: 98ms}  & \adjustbox{stack=cc, margin=0ex 2ex}{50-50\\train-test} & \adjustbox{stack=cc, margin=0ex 2ex}{N/A} & 2019 & \adjustbox{stack=cc, margin=0ex 2ex}{\cite{ramakrishnan2019epileptic}}  \\

        \adjustbox{stack=cc, margin=0ex 2ex}{RF}& \adjustbox{stack=cc, margin=0ex 2ex}{Mean value and peak-to-peak value of wavelet energy obtained after performing PDWC} & \adjustbox{stack=cc, margin=0ex 2ex}{CHB-MIT,\\NICU,\\Pone\_pat,\\Bonn}& \adjustbox{stack=cc, margin=0ex 2ex}{TPR: 99.42\%\\PPV: 99.71\%\\TNR: 99.71\%\\NPV: 99.71\%\\acc: 99.67\%\\F1: 99.54\%}  & \adjustbox{stack=cc, margin=0ex 2ex}{80-20\\train-test} & \adjustbox{stack=cc, margin=0ex 2ex}{4s} & 2021 & \adjustbox{stack=cc, margin=0ex 2ex}{\cite{he2021progressive}}  \\

        \rowcolor{Gray}
       
        \adjustbox{stack=cc, margin=0ex 2ex}{RF}& \adjustbox{stack=cc, margin=0ex 2ex}{Improved sample entropy, phase synchronization index}  & \adjustbox{stack=cc, margin=0ex 2ex}{CHB-MIT\\(2 electrodes)}& \adjustbox{stack=cc, margin=0ex 2ex}{acc: 91.78\%\\sen:  91.27\%\\spe: 93.61\%}  & \adjustbox{stack=cc, margin=0ex 2ex}{10-fold\\CV} & \adjustbox{stack=cc, margin=0ex 2ex}{2s} & 2022 & \adjustbox{stack=cc, margin=0ex 2ex}{\cite{ru2022epilepsy}}\\

        \adjustbox{stack=cc, margin=0ex 2ex}{RF} & \adjustbox{stack=cc, margin=0ex 2ex}{Weighted degree, clustering coefficient, harmonic centrality, modularity, closeness centrality and eigenvector centrality are extracted in 3 networks constructed by PCC, MI and permutation disalignment index} & \adjustbox{stack=cc, margin=0ex 2ex}{CHB-MIT,\\Siena} & \adjustbox{stack=cc, margin=0ex 2ex}{CHB-MIT:\\acc: 97.26\%,\\sen: 96.89\%,\\spe: 97.55\%,\\F1: 97.11\%,\\
        Siena:\\acc: 98.88\%,\\sen: 98.36\%,\\spe:  99.13\%,\\F1: 98.75\%}  & \adjustbox{stack=cc, margin=0ex 2ex}{70-30 train-test,\\ 5-fold CV} & \adjustbox{stack=cc, margin=0ex 2ex}{4s,\\3s overlap} & 2022 & \adjustbox{stack=cc, margin=0ex 2ex}{\cite{xiong2022seizure}}\\ 

        \rowcolor{Gray}

        \adjustbox{stack=cc, margin=0ex 2ex}{RF} & \adjustbox{stack=cc, margin=0ex 2ex}{Power of 6 PSD brain wave bands, vs coherence coefficient} & \adjustbox{stack=cc, margin=0ex 2ex}{TUEP\\(8 channels)}& \adjustbox{stack=cc, margin=0ex 2ex}{Coherence coefficients:\\acc: 90.87\%,\\PSD,\\acc:  95.73\%}  & \adjustbox{stack=cc, margin=0ex 2ex}{70-30\\train-test} & \adjustbox{stack=cc, margin=0ex 2ex}{10s,\\no overlap} & 2023 & \adjustbox{stack=cc, margin=0ex 2ex}{\cite{rohira2023automatic}}  \\
        
        \adjustbox{stack=cc, margin=0ex 2ex}{RF} & \adjustbox{stack=cc, margin=0ex 2ex}{Time and frequency features} & \adjustbox{stack=cc, margin=0ex 2ex}{CHB-MIT,\\private dataset}& \adjustbox{stack=cc, margin=0ex 2ex}{acc: 99.36\%,\\ spe: 82.98\%,\\ sen: 99.41\%,\\ FPR: 0.57 times/h}  & \adjustbox{stack=cc, margin=0ex 2ex}{Leave-5-patient-out} & \adjustbox{stack=cc, margin=0ex 2ex}{5s,\\4s overlap} & 2019 & \adjustbox{stack=cc, margin=0ex 2ex}{\cite{wu2019automatic}}  \\

        \rowcolor{Gray}
        
        \adjustbox{stack=cc, margin=0ex 2ex}{RF} & \adjustbox{stack=cc, margin=0ex 2ex}{Standard deviation, the IQR and energy of each sub-band} & \adjustbox{stack=cc, margin=0ex 2ex}{CHB-MIT}& \adjustbox{stack=cc, margin=0ex 2ex}{acc: 94.04\%,\\sen: 89.5\%,\\pre: 98.4\%}  & \adjustbox{stack=cc, margin=0ex 2ex}{10-fold\\CV} & \adjustbox{stack=cc, margin=0ex 2ex}{10s,\\no overlap} & 2022 & \adjustbox{stack=cc, margin=0ex 2ex}{\cite{misirlis2022pediatric}}  \\
                
        \adjustbox{stack=cc, margin=0ex 2ex}{RF} & \adjustbox{stack=cc, margin=0ex 2ex}{Hjorth parameter, time correlation coefficient matrix, eigenvalues of correlation coefficient matrix, sub-band signal energy, frequency correlation coefficient matrix, fuzzy entropy} & \adjustbox{stack=cc, margin=0ex 2ex}{CHB-MIT}& \adjustbox{stack=cc, margin=0ex 2ex}{acc: 98.03\%,\\spe: 99.04\%,\\sen: 97.02\%}  & \adjustbox{stack=cc, margin=0ex 2ex}{Leave-5-patient-out} & \adjustbox{stack=cc, margin=0ex 2ex}{4s,\\2s overlap} & 2023 & \adjustbox{stack=cc, margin=0ex 2ex}{\cite{dong2023novel}}  \\
        
        %\aboverulesepcolor{Gray}
        \bottomrule

    \end{tabular}
    \end{adjustbox}
    \caption{continued - Feature-based ML methods from systematic review of literature for seizure detection in scalp EEG data}
    \label{ML_2}
\end{table} 

\newpage
\begin{table}[H]
    \footnotesize	
    \centering
    \begin{adjustbox}{width=1\textwidth}
    \begin{tabular}{p{0.2\textwidth}<{\centering}p{0.3\textwidth}<{\centering}ccp{0.15\textwidth}<{\centering}ccp{0.1\textwidth}<{\centering}}
        \toprule 
        \textbf{Classifier} & \textbf{Feature(s)} & \textbf{Dataset(s)} & \textbf{Performance}  & \textbf{Validation} & \textbf{\makecell{Segment\\Length}} & \textbf{Year} & \textbf{References} \\
        \midrule
        \belowrulesepcolor{Gray}
        \rowcolor{Gray}
        
        \adjustbox{stack=cc, margin=0ex 2ex}{LDA\\(classification),\\bagging\\(feature selection)} & \adjustbox{stack=cc, margin=0ex 2ex}{Spectral edge frequencies, spectral edge powers, IQR, MAD, PCC} & \adjustbox{stack=cc, margin=0ex 2ex}{CHB-MIT\\(18 channels)\\AIIMS\\(private)}& \adjustbox{stack=cc, margin=0ex 2ex}{CHB-MIT:\\ acc: 84.83\% \\ FDR: 1.2/hour \\ mean latency: 1.43s}  & \adjustbox{stack=cc, margin=0ex 2ex}{N/A} & \adjustbox{stack=cc, margin=0ex 2ex}{1s,\\no overlap} & 2021  & \adjustbox{stack=cc, margin=0ex 2ex}{\cite{ansari2021patient}}  \\
        
        \adjustbox{stack=cc, margin=0ex 2ex}{LDA} & \adjustbox{stack=cc, margin=0ex 2ex}{Univariate features: kurtosis, mean absolute deviation, interquartile range, and semivariance are calculated after the DWT. Bivariate feature: measure of correlogram} & \adjustbox{stack=cc, margin=0ex 2ex}{CHB-MIT\\(14 patients)}& \adjustbox{stack=cc, margin=0ex 2ex}{sen: 100\%\\FP/Hour: 0.59\\spe: 99.8\%\\ acc: 99.6\%}  & \adjustbox{stack=cc, margin=0ex 2ex}{3-fold\\CV} & \adjustbox{stack=cc, margin=0ex 2ex}{1s,\\no overlap} & 2020 & \adjustbox{stack=cc, margin=0ex 2ex}{\cite{khan2020hybrid}} \\
                
        \rowcolor{Gray}
        
        \adjustbox{stack=cc, margin=0ex 2ex}{XGBoost} & \adjustbox{stack=cc, margin=0ex 2ex}{WAF-based hybrid extracted features, SSA and time-domain features} & \adjustbox{stack=cc, margin=0ex 2ex}{CHB-MIT\\ (18 channels,\\10 patients)}& \adjustbox{stack=cc, margin=0ex 2ex}{acc: 94.46\%\\ sen :88.61\%\\spe: 88.61\%\\precision: 99.81\%\\MCC: 89.54\%\\ kappa: 89.03\%}  & \adjustbox{stack=cc, margin=0ex 2ex}{5-fold\\CV} & \adjustbox{stack=cc, margin=0ex 2ex}{6s,\\no overlap} & 2022 & \adjustbox{stack=cc, margin=0ex 2ex}{\cite{nandini2022efficient}}  \\
        
        \adjustbox{stack=cc, margin=0ex 2ex}{XGBoost} & \adjustbox{stack=cc, margin=0ex 2ex}{Mean, std, signal envelope, kurtosis, skewness, complexity, mobility, TKEO, fractal dimension, band power, sum of relative beta and gamma} & \adjustbox{stack=cc, margin=0ex 2ex}{TUSZ\\(4 channels)}& \adjustbox{stack=cc, margin=0ex 2ex}{sen: 20\%\\FA/24h: 15.59}  & \adjustbox{stack=cc, margin=0ex 2ex}{N/A} & \adjustbox{stack=cc, margin=0ex 2ex}{1s,\\0.5 overlap} & 2020 & \adjustbox{stack=cc, margin=0ex 2ex}{\cite{wei2020epileptic}}  \\

        \rowcolor{Gray}

        \adjustbox{stack=cc, margin=0ex 2ex}{Naive Bayes} & \adjustbox{stack=cc, margin=0ex 2ex}{Relative amplitude, spectral entropy, logarithmic band power, tonal power ratio, 1D local binary pattern, PSD, spectrogram} & \adjustbox{stack=cc, margin=0ex 2ex}{CHB-MIT,\\TUEP} & \adjustbox{stack=cc, margin=0ex 2ex}{TUEP:\\acc: >90\%,\\sen: >85\%,\\spe: >85\%,\\CHBMIT:\\acc: 90\%,\\sen: >92\%,\\spe: >92\%}  & \adjustbox{stack=cc, margin=0ex 2ex}{90-10\\train-test} & \adjustbox{stack=cc, margin=0ex 2ex}{N/A} & 2022 & \adjustbox{stack=cc, margin=0ex 2ex}{\cite{jaffino2022weighted}}\\ 

        \adjustbox{stack=cc, margin=0ex 2ex}{Naive Bayes} & \adjustbox{stack=cc, margin=0ex 2ex}{10 geometric features extracted in each frequency band \(\theta,\beta,\delta,\alpha\)}  & \adjustbox{stack=cc, margin=0ex 2ex}{CHB-MIT} & \adjustbox{stack=cc, margin=0ex 2ex}{acc: 94.54\%}  & \adjustbox{stack=cc, margin=0ex 2ex}{10-fold\\CV} & \adjustbox{stack=cc, margin=0ex 2ex}{20s,\\ 15s overlap} & 2022 & \adjustbox{stack=cc, margin=0ex 2ex}{\cite{wang2022epileptic}}\\ 

        \rowcolor{Gray}

        \adjustbox{stack=cc, margin=0ex 2ex}{Genetic algoritm - Binary Grey Wolf Optimisation} & \adjustbox{stack=cc, margin=0ex 2ex}{
        Std, Shannon entropy, kurtosis, Hjorth parameters, skewness, energy and nonlinear energy, Higuchi fractal dimension, Katz fractal dimension, spectral entropy} & \adjustbox{stack=cc, margin=0ex 2ex}{TUH}& \adjustbox{stack=cc, margin=0ex 2ex}{acc: 85\%}  & \adjustbox{stack=cc, margin=0ex 2ex}{N/A} & \adjustbox{stack=cc, margin=0ex 2ex}{1.8s,\\no overlap} & 2021 & \adjustbox{stack=cc, margin=0ex 2ex}{\cite{davidson2022epileptic}}  \\
        
        \adjustbox{stack=cc, margin=0ex 2ex}{Hidden Markov Model} & \adjustbox{stack=cc, margin=0ex 2ex}{DMD power, sum of 2D PSD, variance, and KFD features} & \adjustbox{stack=cc, margin=0ex 2ex}{CHB-MIT,\\AIIMS}& \adjustbox{stack=cc, margin=0ex 2ex}{average CHB-MIT:\\acc: 99.60\%\\MCC: 0.97\\Kappa: 0.97\\FPR: 0.12\%\\NPV: 99.69\%\\PPV: 98.73\%\\Sen: 96.64\%\\Spe: 99.88\%}  & \adjustbox{stack=cc, margin=0ex 2ex}{N/A} & \adjustbox{stack=cc, margin=0ex 2ex}{5s,\\no overlap}  & 2020 & \adjustbox{stack=cc, margin=0ex 2ex}{ \cite{dash2020multi}}  \\

        \rowcolor{Gray}
        
        \adjustbox{stack=cc, margin=0ex 2ex}{NN} & \adjustbox{stack=cc, margin=0ex 2ex}{AM bandwidth, FM bandwidth, frequency, kurtosis, Hjorth complexity, Hjorth mobility, skewness, spectral centroid, spectral entropy, spectral peak, Spectral power for 8 IMFs}  & \adjustbox{stack=cc, margin=0ex 2ex}{Bonn,\\NSC-HK}& \adjustbox{stack=cc, margin=0ex 2ex}{acc: 98.1\%,\\sen: 98.21 \%,\\spe: 97.65\%}  & \adjustbox{stack=cc, margin=0ex 2ex}{70-30\\train-test} & \adjustbox{stack=cc, margin=0ex 2ex}{N/A} & 2022 & \adjustbox{stack=cc, margin=0ex 2ex}{\cite{kumar2022intelligent}}  \\

        \adjustbox{stack=cc, margin=0ex 2ex}{Multi-layer perceptron} & \adjustbox{stack=cc, margin=0ex 2ex}{Riemannian tangent space map features} & \adjustbox{stack=cc, margin=0ex 2ex}{TUSZ\\(18 channels)} & \adjustbox{stack=cc, margin=0ex 2ex}{acc: 98.94\%,\\Kappa: 0.916} & \adjustbox{stack=cc, margin=0ex 2ex}{N/A} & \adjustbox{stack=cc, margin=0ex 2ex}{6s,\\3s overlap} & 2021 & \adjustbox{stack=cc, margin=0ex 2ex}{\cite{altindics2021detection}}\\

        %\aboverulesepcolor{Gray}
        \bottomrule
              
    \end{tabular}
    \end{adjustbox}
    \caption{continued - Feature-based ML methods from systematic review of literature for seizure detection in scalp EEG data}
    \label{ML_3}
\end{table}

\subsection{DL algorithms for classification}
Whilst feature-based ML requires features as input, DL can uncover patterns and features from different types of data. The input of DL algorithms can be raw or filtered EEG data, any domain representation, or a set of features extracted from EEG signals. DL architectures commonly used for automated seizure detection include artificial neural networks (ANN), CNN and graph machine learning (GML). Different architectures classify an EEG segment based on different properties of the signal. In this section we review these four DL methods. It is worth noting that to leverage the strengths and compensate for the weaknesses of some ML architectures, some studies combine different DL architectures (in parallel or in series). Tables \ref{DL}, \ref{DL2}, \ref{DL3}, \ref{DL4} and \ref{DL5} summarise the DL models encountered in the literature.  

Artificial neural networks are composed of a combination of node layers: an input layer, one or more hidden layers, and an output layer. \citeauthor{sallam2019epilepsy} use the averaged frequency-spectrum values of the \(\delta, \theta, \alpha, \beta\) and \(\gamma\) frequency bands as input to a single layer ANN comprised of 10 hidden neurons. This simple architecture yields an overall accuracy of 93.5\% when trained and tested using the CHB-MIT dataset. 

Convolutional neural networks are the most widely used DL technique in the field of seizure detection: 43 of the 54 deep learning papers included in this study used convolutional layers. The input of a CNN can be 1-dimensional, for example, a single channel EEG trace; 2-dimensional, a tensor where each row represents a channel and each column is a sample; or 3-dimensional, where each segment is represented as a 2D tensor, and samples are concatenated in the third dimension. For instance, \citeauthor{kaziha2020convolutional} and \citeauthor{huang2019automatic} construct a 2D input tensor from 100 and 23 second raw EEG segments from 24 and 21 channels respectively \cite{kaziha2020convolutional, huang2019automatic}. In contrast, \citeauthor{kumar2021convolutional} and \citeauthor{qiu2022lightseizurenet} use 1-dimensional EEG segments as input to a CNN \cite{kumar2021convolutional, qiu2022lightseizurenet}. In \citeauthor{qiu2022lightseizurenet}'s work, the EEG segments of each channel are convoluted separately \cite{qiu2022lightseizurenet}. The input to CNNs is not restricted to the time-domain. Commonly, frequency domain representations are used, or the concatenation of frequency domain features across channels. \citeauthor{sharan2020epileptic} compute the power of 132 different frequency bins for each channel and store it as a 2-dimensional input tensor \cite{sharan2020epileptic}. Similarly, \citeauthor{li2021fft} concatenate the 1-dimensional FFT of each channel to form a 2-dimensional input tensor \cite{li2021fft}. These approaches result in an accuracy of 97.25\% and 98.47\% respectively. \citeauthor{varli2023multiple} combine two different DL algorithms, each with a different input. They input spectrograms and scalograms (absolute value of the CWT of a signal, plotted as a function of time and frequency) into a CNN, and raw EEG traces into a RNN \cite{varli2023multiple}. Combining both DL approaches potentially leverages their respective strengths. Surprisingly, the reported accuracy of 96.23\% is lower than that obtained by other studies using a single CNN.

Unlike ANNs and CNNs, graph machine learning incorporates the spatial information of the electrode placement, which can provide insight on seizure dynamics. Examples of graph machine learning are graph attention networks (GAT), and graph convolution networks (GCN). \citeauthor{zhao2021graph} use a GAT with the CHB-MIT dataset, and obtain an accuracy, sensitivity and specificity of 98.89\%, 97.10\% and 99.63\%, respectively \cite{zhao2021graph}.

Some methods combine ML and DL. For example, \citeauthor{dalal2022statistical} use statistical features like kurtosis, variance and skew as the input to multiple CNNs \cite{dalal2022statistical}. The output of the CNNs are concatenated and used as input to a LSTM (a type of RNN) for binary classification. They achieve an accuracy of 94.6\%, recall of 97.15\%, and precision of 95.78\%.

There are instances where increasing complexity of the respective DL algorithm does not lead to an increase in accuracy. For example, \citeauthor{yan2019automated} trained four different CNNs \cite{yan2019automated}. They observed that the weights of the CNN with the highest complexity (4 convolution blocks/10 layers) did not converge during training. Consequently, the CNN fails to perform the classification task. The authors suggest that after using four convolution blocks the extracted spectrographic features are not seizure specific anymore \cite{yan2019automated}.\\

% DL tables
\newpage
\begin{table}[H]
    \centering
    \begin{adjustbox}{width=\textwidth}
    \scriptsize	
    \begin{tabular}{>{\centering}p{0.25\textwidth}cccccp{0.1\textwidth}<{\centering}}
        \toprule 
        \textbf{Classifier}  & \textbf{Dataset(s)} & \textbf{Performance}  & \textbf{Validation} & \textbf{\makecell{Segment\\Length}} & \textbf{Year} & \textbf{References} \\
        \midrule
        \belowrulesepcolor{Gray}
        \rowcolor{Gray}

        \adjustbox{stack=cc, margin=0ex 2ex}{CNN}  & \adjustbox{stack=cc, margin=0ex 2ex}{CHB-MIT} & \adjustbox{stack=cc, margin=0ex 2ex}{sen: 97.25\%, \\ spe: 97.25\%, \\ acc: 97.25\%}  & \adjustbox{stack=cc, margin=0ex 2ex}{10-fold\\CV} & \adjustbox{stack=cc, margin=0ex 2ex}{3s} & \adjustbox{stack=cc, margin=0ex 2ex}{2020} & \adjustbox{stack=cc, margin=0ex 2ex}{\cite{sharan2020epileptic}}  \\

        \adjustbox{stack=cc, margin=0ex 2ex}{CNN} & \adjustbox{stack=cc, margin=0ex 2ex}{CHB-MIT} & \adjustbox{stack=cc, margin=0ex 2ex}{acc: 96.74\% \\ spe: 100\% \\ sen: 82.35\%}  & \adjustbox{stack=cc, margin=0ex 2ex}{5-fold\\CV} & \adjustbox{stack=cc, margin=0ex 2ex}{100s} & \adjustbox{stack=cc, margin=0ex 2ex}{2020} & \adjustbox{stack=cc, margin=0ex 2ex}{\cite{kaziha2020convolutional}}  \\

        \rowcolor{Gray}

        \adjustbox{stack=cc, margin=0ex 2ex}{CNN} & \adjustbox{stack=cc, margin=0ex 2ex}{CHB-MIT} & \adjustbox{stack=cc, margin=0ex 2ex}{acc: 87.4\% \\ sen:88.10\% \\spe:87.10\%\\F1:87.40\%\\ pre: 86.98\%}  & \adjustbox{stack=cc, margin=0ex 2ex}{10-fold\\CV} & 
        \adjustbox{stack=cc, margin=0ex 2ex}{8s} & 
        \adjustbox{stack=cc, margin=0ex 2ex}{2021} & \adjustbox{stack=cc, margin=0ex 2ex}{\cite{kumar2021convolutional}}  \\

        \adjustbox{stack=cc, margin=0ex 2ex}{CNN} & \adjustbox{stack=cc, margin=0ex 2ex}{CHB-MIT,\\Bonn} & \adjustbox{stack=cc, margin=0ex 2ex}{acc: 96.69\% \\ sen: 96.19\% \\ spe: 97.08\%}  & \adjustbox{stack=cc, margin=0ex 2ex}{k-fold\\CV} & \adjustbox{stack=cc, margin=0ex 2ex}{2s} & 
        \adjustbox{stack=cc, margin=0ex 2ex}{2023} & \adjustbox{stack=cc, margin=0ex 2ex}{\cite{cimr2023automatic}}  \\

        \rowcolor{Gray}

        \adjustbox{stack=cc, margin=0ex 2ex}{CNN} & \adjustbox{stack=cc, margin=0ex 2ex}{CHB-MIT,\\Bonn} & \adjustbox{stack=cc, margin=0ex 2ex}{acc: 98.80\% \\ sen: \>98\% \\ spe: \>98\%}  & \adjustbox{stack=cc, margin=0ex 2ex}{10-fold\\CV} & \adjustbox{stack=cc, margin=0ex 2ex}{N/A} & \adjustbox{stack=cc, margin=0ex 2ex}{2021} & \adjustbox{stack=cc, margin=0ex 2ex}{\cite{ramakrishnan2021seizure}}  \\

        \adjustbox{stack=cc, margin=0ex 2ex}{CNN + MIDS,\\CNN + data augmentation} & \adjustbox{stack=cc, margin=0ex 2ex}{CHB-MIT} & \adjustbox{stack=cc, margin=0ex 2ex}{CNN+MIDS: \\sen: 74.08\% \\spe: 92.46\%\\CNN+Data aug:\\sen: \> 72.11\% \\spe: 95.89\%\\ }  & \adjustbox{stack=cc, margin=0ex 2ex}{LOO} & \adjustbox{stack=cc, margin=0ex 2ex}{5s} & \adjustbox{stack=cc, margin=0ex 2ex}{2019} & \adjustbox{stack=cc, margin=0ex 2ex}{\cite{
        wei2019automatic}}  \\

        \rowcolor{Gray}

        \adjustbox{stack=cc, margin=0ex 2ex}{CNN aided\\factor graph} & \adjustbox{stack=cc, margin=0ex 2ex}{CHB-MIT} & \adjustbox{stack=cc, margin=0ex 2ex}{AUC-ROC: 90.23\% \\ AUC-PR: 76.77\%\\ F1: 90.42\%\\ }  & \adjustbox{stack=cc, margin=0ex 2ex}{ 6 fold, leave\\4 patients out} & \adjustbox{stack=cc, margin=0ex 2ex}{4s} & \adjustbox{stack=cc, margin=0ex 2ex}{2021} & \adjustbox{stack=cc, margin=0ex 2ex}{\cite{salafian2021efficient}}  \\

        \adjustbox{stack=cc, margin=0ex 2ex}{CNN aided\\factor graph}  & \adjustbox{stack=cc, margin=0ex 2ex}{CHB-MIT} & \adjustbox{stack=cc, margin=0ex 2ex}{AUC-ROC: 83.8\% \\ AUC-PR: 50.38\%\\ F1: 93.42\%}  & \adjustbox{stack=cc, margin=0ex 2ex}{6 fold, leave \\4 patients out} & \adjustbox{stack=cc, margin=0ex 2ex}{4s and\\ 32s} & \adjustbox{stack=cc, margin=0ex 2ex}{2022} & \adjustbox{stack=cc, margin=0ex 2ex}{\cite{salafian2022cnn}}  \\

        \rowcolor{Gray}

        \adjustbox{stack=cc, margin=0ex 2ex}{Attention-based\\CNN-BiRNN} & \adjustbox{stack=cc, margin=0ex 2ex}{CHB-MIT} & \adjustbox{stack=cc, margin=0ex 2ex}{No missing channel: \\ 
        spe: 93.94\% \\ sen: 92.88\%\\2 missing channels:\\ spe: \>90\%\\ sen: \>95\%}  & \adjustbox{stack=cc, margin=0ex 2ex}{10-fold\\CV} & \adjustbox{stack=cc, margin=0ex 2ex}{23s} & 2019 & \adjustbox{stack=cc, margin=0ex 2ex}{\cite{huang2019automatic}}  \\

        \adjustbox{stack=cc, margin=0ex 2ex} {Medium weight\\deep CNN}  & \adjustbox{stack=cc, margin=0ex 2ex}{CHB-MIT} & \adjustbox{stack=cc, margin=0ex 2ex}{acc: 96\%\\}  & \adjustbox{stack=cc, margin=0ex 2ex}{10-fold\\CV} & \adjustbox{stack=cc, margin=0ex 2ex}{300ms,\\20ms overlap} &  \adjustbox{stack=cc, margin=0ex 2ex}{2022} & \adjustbox{stack=cc, margin=0ex 2ex}{\cite{nemati2022medium}}  \\

        \rowcolor{Gray}       
        \adjustbox{stack=cc, margin=0ex 2ex}{Hybrid Probabilistic\\Graphical Model\\CNN (PGM-CNN)} & \adjustbox{stack=cc, margin=0ex 2ex}{CHB-MIT, \\Johns Hopkins\\ Hospital (JHH)} & \adjustbox{stack=cc, margin=0ex 2ex}{TPR: 0.61\\ FPR: 0.0067\\AUC:0.8\\F1: 0.67\\pre: 0.83\\}  & \adjustbox{stack=cc, margin=0ex 2ex}{5-fold\\CV} & \adjustbox{stack=cc, margin=0ex 2ex}{1s} & \adjustbox{stack=cc, margin=0ex 2ex}{2019} & \adjustbox{stack=cc, margin=0ex 2ex}{\cite{craley2019integrating}}  \\

        \adjustbox{stack=cc, margin=0ex 2ex}{GCN} & \adjustbox{stack=cc, margin=0ex 2ex}{CHB-MIT} & \adjustbox{stack=cc, margin=0ex 2ex}{acc: 98.35\%}  & \adjustbox{stack=cc, margin=0ex 2ex}{10-fold\\CV} & \adjustbox{stack=cc, margin=0ex 2ex}{60s} & \adjustbox{stack=cc, margin=0ex 2ex}{2020} & \adjustbox{stack=cc, margin=0ex 2ex}{\cite{chen2020epilepsy}}  \\

        \rowcolor{Gray}
        \adjustbox{stack=cc, margin=0ex 2ex}{3D-CNN} & \adjustbox{stack=cc, margin=0ex 2ex}{CHB-MIT,\\TUH} & \adjustbox{stack=cc, margin=0ex 2ex}{CHB-MIT:\\acc: 94.36\% \\ rec: 95.57\% \\ TUH:\\acc: 92.26\% \\ rec: 93.86\%}  & \adjustbox{stack=cc, margin=0ex 2ex}{N/A} & \adjustbox{stack=cc, margin=0ex 2ex}{2s}  & \adjustbox{stack=cc, margin=0ex 2ex}{2021} & \adjustbox{stack=cc, margin=0ex 2ex}{\cite{sun2022automatic}}  \\

        \aboverulesepcolor{Gray}
        \bottomrule
              
    \end{tabular}
    \end{adjustbox}
    \caption{DL methods from systematic review of literature for seizure detection in scalp EEG data}
    \label{DL}
   
\end{table} 

\newpage
\begin{table}[H]
    \centering
    \begin{adjustbox}{width=\textwidth}
    \scriptsize	
    \begin{tabular}{>{\centering}p{0.25\textwidth}cccccp{0.1\textwidth}<{\centering}}
        \toprule 
        \textbf{Classifier}  & \textbf{Dataset(s)} & \textbf{Performance}  & \textbf{Validation} & \textbf{\makecell{Segment\\Length}} & \textbf{Year} & \textbf{References} \\
        \midrule
        \belowrulesepcolor{Gray}

        \rowcolor{Gray}
        \adjustbox{stack=cc, margin=0ex 2ex}{Deep CNN} & \adjustbox{stack=cc, margin=0ex 2ex}{CHB-MIT,\\Bonn}& \adjustbox{stack=cc, margin=0ex 2ex}{acc: 91.82\%\\ sen: 91.93\% \\ F1: 95.73\% \\ FRP: 0.005/hour}  & \adjustbox{stack=cc, margin=0ex 2ex}{three-way\\holdout} & \adjustbox{stack=cc, margin=0ex 2ex}{5s\\no overlap} & \adjustbox{stack=cc, margin=0ex 2ex}{2021} & \adjustbox{stack=cc, margin=0ex 2ex}{\cite{tallon2022effective}}  \\

        \adjustbox{stack=cc, margin=0ex 2ex}{CNN}  & \adjustbox{stack=cc, margin=0ex 2ex}{CHB-MIT}& \adjustbox{stack=cc, margin=0ex 2ex}{acc: 93.4\%}  & \adjustbox{stack=cc, margin=0ex 2ex}{6-fold\\CV} & \adjustbox{stack=cc, margin=0ex 2ex}{1s} & \adjustbox{stack=cc, margin=0ex 2ex}{2022} & \adjustbox{stack=cc, margin=0ex 2ex}{\cite{sukaria2022epileptic}}  \\

        \rowcolor{Gray}
        \adjustbox{stack=cc, margin=0ex 2ex}{2D-PCANet\\(feature extraction)\\ SVM\\(classification)}  & \adjustbox{stack=cc, margin=0ex 2ex}{CHB-MIT\\Bonn}& \adjustbox{stack=cc, margin=0ex 2ex}{acc: 98.47\% \\ sen: 98.28\% \\ spe: 98.50\% \\}  & \adjustbox{stack=cc, margin=0ex 2ex}{10-fold\\CV} & \adjustbox{stack=cc, margin=0ex 2ex}{1s} & \adjustbox{stack=cc, margin=0ex 2ex}{2021} & \adjustbox{stack=cc, margin=0ex 2ex}{\cite{li2021fft}}  \\

        \adjustbox{stack=cc, margin=0ex 2ex}{GBDT, attention-based\\ CNN-BiRNN, FC layer\\ for classification} & \adjustbox{stack=cc, margin=0ex 2ex}{CHB-MIT}& \adjustbox{stack=cc, margin=0ex 2ex}{acc: 97.56\% \\ sen: 90.97\% \\ spe: 91.93\% \\}  & \adjustbox{stack=cc, margin=0ex 2ex}{train-val-test\\70-15-15} & \adjustbox{stack=cc, margin=0ex 2ex}{20s} & \adjustbox{stack=cc, margin=0ex 2ex}{2021} & \adjustbox{stack=cc, margin=0ex 2ex}{\cite{huang2021feature}}  \\

        \rowcolor{Gray}
        \adjustbox{stack=cc, margin=0ex 2ex}{CNN, LSTM} & \adjustbox{stack=cc, margin=0ex 2ex}{CHB-MIT\\(22 patients, 8 channels)}& \adjustbox{stack=cc, margin=0ex 2ex}{acc: 94.6\%\\ rec: 97.15\% \\pre: 95.78\%}  & \adjustbox{stack=cc, margin=0ex 2ex}{10-fold\\CV} & \adjustbox{stack=cc, margin=0ex 2ex}{N/A} & \adjustbox{stack=cc, margin=0ex 2ex}{2022}  & \adjustbox{stack=cc, margin=0ex 2ex}{\cite{dalal2022statistical}}  \\

        \adjustbox{stack=cc, margin=0ex 2ex}{1D CNN}& \adjustbox{stack=cc, margin=0ex 2ex}{CHB-MIT\\(21 channels)}& \adjustbox{stack=cc, margin=0ex 2ex}{acc: 97.09\% \\sen: 96.49\% \\spe: 97.09\% \\}  & \adjustbox{stack=cc, margin=0ex 2ex}{10-fold\\CV} & \adjustbox{stack=cc, margin=0ex 2ex}{2s, 1s\\overlap}  & \adjustbox{stack=cc, margin=0ex 2ex}{2022} & \adjustbox{stack=cc, margin=0ex 2ex}{\cite{qiu2022lightseizurenet}}  \\

        \rowcolor{Gray}
        \adjustbox{stack=cc, margin=0ex 2ex}{ResNet-based} & \adjustbox{stack=cc, margin=0ex 2ex}{TUSZ\\(20 channels)}& \adjustbox{stack=cc, margin=0ex 2ex}{acc: 69\%\\segment level,\\acc: 61.67\%}  & \adjustbox{stack=cc, margin=0ex 2ex}{3-fold\\CV} & \adjustbox{stack=cc, margin=0ex 2ex}{1s, 0.75s\\overlap} & \adjustbox{stack=cc, margin=0ex 2ex}{2022} & \adjustbox{stack=cc, margin=0ex 2ex}{\cite{tiwary2022deep}}  \\

        \adjustbox{stack=cc, margin=0ex 2ex}{Attention\\based CNN}  & \adjustbox{stack=cc, margin=0ex 2ex}{TUSZ\\(14 subjects)}& \adjustbox{stack=cc, margin=0ex 2ex}{sen: 97.4\%\\spe:  88.1\%\\acc: 80.5\%}  & \adjustbox{stack=cc, margin=0ex 2ex}{LOO, and\\14-fold\\CV} & \adjustbox{stack=cc, margin=0ex 2ex}{1s, 0.5s\\overlap} & \adjustbox{stack=cc, margin=0ex 2ex}{2022} & \adjustbox{stack=cc, margin=0ex 2ex}{\cite{zhang2020adversarial}}  \\

        \rowcolor{Gray}
        \adjustbox{stack=cc, margin=0ex 2ex}{U-net(feature\\extraction),\\LSTM\\(classification)} & \adjustbox{stack=cc, margin=0ex 2ex}{TUSZ \\(16 channels)}& \adjustbox{stack=cc, margin=0ex 2ex}{sen: 12.37\\Fas/24hr: 1.44\\TAES score: 2.46\\}  & \adjustbox{stack=cc, margin=0ex 2ex}{10-fold\\CV} & \adjustbox{stack=cc, margin=0ex 2ex}{20s} & \adjustbox{stack=cc, margin=0ex 2ex}{2020} & \adjustbox{stack=cc, margin=0ex 2ex}{\cite{chatzichristos2020epileptic}}  \\

        \adjustbox{stack=cc, margin=0ex 2ex}{CNN-SVM} & \adjustbox{stack=cc, margin=0ex 2ex}{CHB-MIT}& \adjustbox{stack=cc, margin=0ex 2ex}{acc: 98.31\%}  & \adjustbox{stack=cc, margin=0ex 2ex}{train-test-val:\\70/15/15	} & \adjustbox{stack=cc, margin=0ex 2ex}{N/A} & \adjustbox{stack=cc, margin=0ex 2ex}{2022} & \adjustbox{stack=cc, margin=0ex 2ex}{\cite{ipek2022towards}}  \\
        
        \rowcolor{Gray}
        \adjustbox{stack=cc, margin=0ex 2ex}{CNNs, FC layer} & \adjustbox{stack=cc, margin=0ex 2ex}{CHB-MIT\\(remove\\patient 12, 21\\channels),\\TUSZ\\(28 patients)}& \adjustbox{stack=cc, margin=0ex 2ex}{CHB-MIT:\\acc: 96.17\%\\sen: 56.83\%\\spe: 96.97\%\\F1: 38.26\%\\TUSZ:\\acc: 67.68\%\\sen:  59.21\%\\spe: 75.30\%\\F1:  47.55\%\\}  & \adjustbox{stack=cc, margin=0ex 2ex}{5-fold\\CV} & \adjustbox{stack=cc, margin=0ex 2ex}{4s, 1s\\overlap}  & \adjustbox{stack=cc, margin=0ex 2ex}{2021} & \adjustbox{stack=cc, margin=0ex 2ex}{\cite{thuwajit2021eegwavenet}}\\

        \adjustbox{stack=cc, margin=0ex 2ex}{CNN} & \adjustbox{stack=cc, margin=0ex 2ex}{CHB-MIT\\(8 channels,\\16 patients)}& \adjustbox{stack=cc, margin=0ex 2ex}{acc: 97.57\%\\sen: 98.90\%\\FPR:2.13\%\\delay:10.46s}  & \adjustbox{stack=cc, margin=0ex 2ex}{LOO} & \adjustbox{stack=cc, margin=0ex 2ex}{5s,\\1s overlap} & \adjustbox{stack=cc, margin=0ex 2ex}{2023} & \adjustbox{stack=cc, margin=0ex 2ex}{\cite{shen2023real}}  \\

        \rowcolor{Gray}
        \adjustbox{stack=cc, margin=0ex 2ex}{CNN}  & \adjustbox{stack=cc, margin=0ex 2ex}{NYP-WC,\\CHB-MIT}& \adjustbox{stack=cc, margin=0ex 2ex}{-}  & \adjustbox{stack=cc, margin=0ex 2ex}{5-fold\\CV} & \adjustbox{stack=cc, margin=0ex 2ex}{120s,\\119 overlap} & \adjustbox{stack=cc, margin=0ex 2ex}{2019} & \adjustbox{stack=cc, margin=0ex 2ex}{\cite{yan2019automated}}  \\

        \adjustbox{stack=cc, margin=0ex 2ex}{CNNs with an\\attention\\mechanism}  & \adjustbox{stack=cc, margin=0ex 2ex}{TUH}& \adjustbox{stack=cc, margin=0ex 2ex}{acc: 86\%,\\F1: 81\%}  & \adjustbox{stack=cc, margin=0ex 2ex}{LOO} & \adjustbox{stack=cc, margin=0ex 2ex}{3s,\\no overlap} & \adjustbox{stack=cc, margin=0ex 2ex}{2023} & \adjustbox{stack=cc, margin=0ex 2ex}{\cite{einizade2023explainable}}  \\
        %\aboverulesepcolor{Gray}
        \bottomrule
              
    \end{tabular}
    \end{adjustbox}
    \caption{continued - DL methods from systematic review of literature for seizure detection in scalp EEG data}
    \label{DL2}
    \vspace*{1 cm}
\end{table} 

\newpage
\begin{table}[H]
    \centering
    \begin{adjustbox}{width=\textwidth}
    \scriptsize	
    \begin{tabular}{>{\centering}p{0.25\textwidth}cccccp{0.1\textwidth}<{\centering}}
        \toprule 
        \textbf{Classifier}  & \textbf{Dataset(s)} & \textbf{Performance}  & \textbf{Validation} & \textbf{\makecell{Segment\\Length}} & \textbf{Year} & \textbf{References} \\
        \midrule
        \belowrulesepcolor{Gray}
        \rowcolor{Gray}

        \adjustbox{stack=cc, margin=0ex 2ex}{CNN vs Xception}  & \adjustbox{stack=cc, margin=0ex 2ex}{CHB-MIT}& \adjustbox{stack=cc, margin=0ex 2ex}{CNN:\\acc: 98.47\%,\\pre: 99.79\%,\\ rec:  98.93\%,\\ F1:  98.51\%\\Xception:\\acc: 95.52\%,\\pre: 99.93\%,\\ rec:  98.63\%,\\ F1:  97.05\%}  & \adjustbox{stack=cc, margin=0ex 2ex}{CV} & \adjustbox{stack=cc, margin=0ex 2ex}{N/A} & \adjustbox{stack=cc, margin=0ex 2ex}{2022} & \adjustbox{stack=cc, margin=0ex 2ex}{\cite{sagga2022epileptic}}  \\
     
        \adjustbox{stack=cc, margin=0ex 2ex}{ResNest18} & \adjustbox{stack=cc, margin=0ex 2ex}{TUSZ}& \adjustbox{stack=cc, margin=0ex 2ex}{sen: 42.05\%,\\FAR/day : 5.78}  & \adjustbox{stack=cc, margin=0ex 2ex}{CV} & \adjustbox{stack=cc, margin=0ex 2ex}{250 samples}  & \adjustbox{stack=cc, margin=0ex 2ex}{2021} & \adjustbox{stack=cc, margin=0ex 2ex}{\cite{khalkhali2021low}}  \\

        \rowcolor{Gray}
        \adjustbox{stack=cc, margin=0ex 2ex}{Multi-fuse reduced deep CNN (MF-RDCNN)} & \adjustbox{stack=cc, margin=0ex 2ex}{Bonn,\\CHB-MIT,\\Neurology Sleep\\Centre Delhi }& \adjustbox{stack=cc, margin=0ex 2ex}{CHB-MIT:\\acc: 99.29\%,\\sen: 99.29\%,\\spe: 99.86\%,\\FPR: 0.71\%}  & \adjustbox{stack=cc, margin=0ex 2ex}{40-40-20\\train-test-val} & \adjustbox{stack=cc, margin=0ex 2ex}{N/A} & \adjustbox{stack=cc, margin=0ex 2ex}{2022}  & \adjustbox{stack=cc, margin=0ex 2ex}{\cite{rout2022efficient}}  \\

        \adjustbox{stack=cc, margin=0ex 2ex}{Multilayer deep convolutional neural network (MDCNN)} & \adjustbox{stack=cc, margin=0ex 2ex}{CHB-MIT (18 subjects, all\\of which have 23 channels)} & \adjustbox{stack=cc, margin=0ex 2ex}{acc: 71.60\%}  & \adjustbox{stack=cc, margin=0ex 2ex}{LOO} & \adjustbox{stack=cc, margin=0ex 2ex}{1s,\\0.5 overlap} & \adjustbox{stack=cc, margin=0ex 2ex}{2021}  & \adjustbox{stack=cc, margin=0ex 2ex}{\cite{dang2021studying}}  \\

        \rowcolor{Gray}
        \adjustbox{stack=cc, margin=0ex 2ex}{CNN using adversarial network methods}  & \adjustbox{stack=cc, margin=0ex 2ex}{CHB-MIT\\(18 channels)}& \adjustbox{stack=cc, margin=0ex 2ex}{acc: 91.71\%,\\sen: 91.09\%,\\spe: 94.73\%,\\FPR: 0.58/hr,\\latency: 4.45s}  & \adjustbox{stack=cc, margin=0ex 2ex}{LOO} & \adjustbox{stack=cc, margin=0ex 2ex}{4s,\\no overlap} & \adjustbox{stack=cc, margin=0ex 2ex}{2022} & \adjustbox{stack=cc, margin=0ex 2ex}{\cite{nasiri2021generalizable}}  \\

        \adjustbox{stack=cc, margin=0ex 2ex}{CNN and RNN}  & \adjustbox{stack=cc, margin=0ex 2ex}{CHB-MIT,\\Bonn,\\Bern-Barcelona}& \adjustbox{stack=cc, margin=0ex 2ex}{acc: 96.23\%}  & \adjustbox{stack=cc, margin=0ex 2ex}{8-fold\\CV} & \adjustbox{stack=cc, margin=0ex 2ex}{N/A} & \adjustbox{stack=cc, margin=0ex 2ex}{2023} & \adjustbox{stack=cc, margin=0ex 2ex}{\cite{varli2023multiple}}  \\

        \rowcolor{Gray}
        \adjustbox{stack=cc, margin=0ex 2ex}{AttVGGNet-RC}  & \adjustbox{stack=cc, margin=0ex 2ex}{CHB-MIT (23\\channels, remove\\patient 12)}& \adjustbox{stack=cc, margin=0ex 2ex}{sen: 93.84 \(\pm\) 0.63\%,\\spe: 95.84 \(\pm\) 0.74\%,\\acc: 95.12 \(\pm\) 0.20\%}  & \adjustbox{stack=cc, margin=0ex 2ex}{10-fold\\CV} & \adjustbox{stack=cc, margin=0ex 2ex}{1s,\\no overlap} & \adjustbox{stack=cc, margin=0ex 2ex}{2020} & \adjustbox{stack=cc, margin=0ex 2ex}{\cite{zhang2020automatic}}  \\

        \adjustbox{stack=cc, margin=0ex 2ex}{CNN (feature extraction),\\and ANN, LR, RF, SVM,\\GB, k-NN, SGD,\\Ensembles (classification)}  & \adjustbox{stack=cc, margin=0ex 2ex}{CHB-MIT,\\Bonn}& \adjustbox{stack=cc, margin=0ex 2ex}{ANN: 94.4\%,\\LR: 91.7\%,\\RF: 92.4\%,\\SVM: 95.7\%,\\GB: 94.6\%,\\k-NN: 96.8\%,\\SGD: 87\%,\\Ensembles: 97\%}  & \adjustbox{stack=cc, margin=0ex 2ex}{10-fold\\CV} & \adjustbox{stack=cc, margin=0ex 2ex}{5s,\\no overlap} & \adjustbox{stack=cc, margin=0ex 2ex}{2022} & \adjustbox{stack=cc, margin=0ex 2ex}{\cite{hassan2022epileptic}}  \\

        \rowcolor{Gray}
        \adjustbox{stack=cc, margin=0ex 2ex}{ANN} & \adjustbox{stack=cc, margin=0ex 2ex}{CHB-MIT}& \adjustbox{stack=cc, margin=0ex 2ex}{N/A}  & \adjustbox{stack=cc, margin=0ex 2ex}{N/A} & \adjustbox{stack=cc, margin=0ex 2ex}{100s,\\no overlap} & \adjustbox{stack=cc, margin=0ex 2ex}{2019} & \adjustbox{stack=cc, margin=0ex 2ex}{\cite{sallam2019epilepsy}}  \\

        \adjustbox{stack=cc, margin=0ex 2ex}{Asymmetrical Back Propagation Neural Network (ABPN)}  & \adjustbox{stack=cc, margin=0ex 2ex}{CHB-MIT} & \adjustbox{stack=cc, margin=0ex 2ex}{sen: 96.32\%,\\spe: 95.12\%,\\acc: 98.36\%}  & \adjustbox{stack=cc, margin=0ex 2ex}{N/A} & \adjustbox{stack=cc, margin=0ex 2ex}{N/A} & \adjustbox{stack=cc, margin=0ex 2ex}{2021} & \adjustbox{stack=cc, margin=0ex 2ex}{\cite{poorani2021seizure}}  \\

        \rowcolor{Gray}
        \adjustbox{stack=cc, margin=0ex 2ex}{BERT (LLM)} & \adjustbox{stack=cc, margin=0ex 2ex}{TUSZ}& \adjustbox{stack=cc, margin=0ex 2ex}{acc:  \(\sim\)77\%}  & \adjustbox{stack=cc, margin=0ex 2ex}{N/A} & \adjustbox{stack=cc, margin=0ex 2ex}{1s,\\no overlap}  & \adjustbox{stack=cc, margin=0ex 2ex}{2022} & \adjustbox{stack=cc, margin=0ex 2ex}{\cite{davidson2022seizure}}  \\

        \adjustbox{stack=cc, margin=0ex 2ex}{CNViT (Convolutional Vision Transformer) that first uses multi-layer convolution to extract features, and then adopts transformer blocks} & \adjustbox{stack=cc, margin=0ex 2ex}{CHB-MIT}& \adjustbox{stack=cc, margin=0ex 2ex}{sen: 96.71\%,\\spe: 97.23\%,\\acc: 97.15\%,\\AUC: 99.54\%}  & \adjustbox{stack=cc, margin=0ex 2ex}{N/A} & \adjustbox{stack=cc, margin=0ex 2ex}{2s,\\no overlap} & \adjustbox{stack=cc, margin=0ex 2ex}{2022} & \adjustbox{stack=cc, margin=0ex 2ex}{\cite{ke2022convolutional}}  \\

        \rowcolor{Gray}
        \adjustbox{stack=cc, margin=0ex 2ex}{Graph isomorphism network (GIN)} & \adjustbox{stack=cc, margin=0ex 2ex}{CHB-MIT}& \adjustbox{stack=cc, margin=0ex 2ex}{acc: 96.2\%,\\sen:  95.4\%,\\spe: 97.0\%}  & \adjustbox{stack=cc, margin=0ex 2ex}{10-fold\\CV} & \adjustbox{stack=cc, margin=0ex 2ex}{20s,\\no overlap} & \adjustbox{stack=cc, margin=0ex 2ex}{2022} & \adjustbox{stack=cc, margin=0ex 2ex}{\cite{tao2022seizure}}  \\

        \aboverulesepcolor{Gray}
        \bottomrule
              
    \end{tabular}
    \end{adjustbox}
    \caption{continued - DL methods from systematic review of literature for seizure detection in scalp EEG data}
    \label{DL3}
\end{table} 

%TO DO%
\newpage
\begin{table}[H]
    \centering
    \begin{adjustbox}{width=\textwidth}
    \scriptsize	
    \begin{tabular}{>{\centering}p{0.25\textwidth}cccccp{0.1\textwidth}<{\centering}}
        \toprule 
        \textbf{Classifier}  & \textbf{Dataset(s)} & \textbf{Performance}  & \textbf{Validation} & \textbf{\makecell{Segment\\Length}} & \textbf{Year} & \textbf{References} \\
        \midrule
        \belowrulesepcolor{Gray}
        \rowcolor{Gray}

        \adjustbox{stack=cc, margin=0ex 2ex}{Graph-generative neural network (GGN)} & \adjustbox{stack=cc, margin=0ex 2ex}{TUH}& \adjustbox{stack=cc, margin=0ex 2ex}{acc: 91\%}  & \adjustbox{stack=cc, margin=0ex 2ex}{70-30\\train-test} & \adjustbox{stack=cc, margin=0ex 2ex}{5s,\\no overlap} & \adjustbox{stack=cc, margin=0ex 2ex}{2022} & \adjustbox{stack=cc, margin=0ex 2ex}{\cite{li2022graph}}  \\

        \adjustbox{stack=cc, margin=0ex 2ex}{GAT and BiLSTM}  & \adjustbox{stack=cc, margin=0ex 2ex}{CHB-MIT,\\TUH}& \adjustbox{stack=cc, margin=0ex 2ex}{CHB-MIT:\\acc: 98.52\%,\\spe: 94.34\%,\\sen: 97.75\%,\\TUH:\\acc: 98.02\%,\\spe: 99.06\%,\\sen: 97.7\%}  & \adjustbox{stack=cc, margin=0ex 2ex}{5-fold\\CV} & \adjustbox{stack=cc, margin=0ex 2ex}{1s,\\0.5s overlap} & \adjustbox{stack=cc, margin=0ex 2ex}{2022} & \adjustbox{stack=cc, margin=0ex 2ex}{\cite{he2022spatial}}  \\

        \rowcolor{Gray}
        \adjustbox{stack=cc, margin=0ex 2ex}{AE (feature extraction), RF (classification)}  & \adjustbox{stack=cc, margin=0ex 2ex}{Siena}& \adjustbox{stack=cc, margin=0ex 2ex}{F1 ictal: 91\%,\\F1 non-ictal: 90.1\%}  & \adjustbox{stack=cc, margin=0ex 2ex}{Leave-2-out} & \adjustbox{stack=cc, margin=0ex 2ex}{6s, 1s overlap} & \adjustbox{stack=cc, margin=0ex 2ex}{2021} & \adjustbox{stack=cc, margin=0ex 2ex}{\cite{ferariu2021using}}  \\

        \adjustbox{stack=cc, margin=0ex 2ex}{Deep convolutional Autoencoder Bi-LSTM}  & \adjustbox{stack=cc, margin=0ex 2ex}{CHB-MIT}& \adjustbox{stack=cc, margin=0ex 2ex}{sen: 99.7\%,\\ acc: 99.8\%,\\spe: 99.9\%,\\precision: 99.9\%,\\F1: 99.6\%}  & \adjustbox{stack=cc, margin=0ex 2ex}{10-fold\\CV} & \adjustbox{stack=cc, margin=0ex 2ex}{4s, no overlap} & \adjustbox{stack=cc, margin=0ex 2ex}{2023} & \adjustbox{stack=cc, margin=0ex 2ex}{\cite{mir2023deep}}  \\

        \rowcolor{Gray}
        \adjustbox{stack=cc, margin=0ex 2ex}{Deep Stacked AE}  & \adjustbox{stack=cc, margin=0ex 2ex}{CHB-MIT,\\TUEP}& \adjustbox{stack=cc, margin=0ex 2ex}{TUEP:\\acc: 91.5\%,\\sen: 85.2\%,\\spe: 86.0\%,\\CHBMIT:\\acc: 91.4\%,\\sen: 85.5\%,\\spe: 85.3\%}  & \adjustbox{stack=cc, margin=0ex 2ex}{90-10\\train-test} & \adjustbox{stack=cc, margin=0ex 2ex}{N/A} & \adjustbox{stack=cc, margin=0ex 2ex}{2022} & \adjustbox{stack=cc, margin=0ex 2ex}{\cite{jose2021adaptive}}  \\

        \adjustbox{stack=cc, margin=0ex 2ex}{AE}  & \adjustbox{stack=cc, margin=0ex 2ex}{CHB-MIT\\(13 patients,\\at least 4 seizure/patient)}& \adjustbox{stack=cc, margin=0ex 2ex}{Sen: 0.86\%,\\FPR/h: 0.08\%}  & \adjustbox{stack=cc, margin=0ex 2ex}{LOO} & \adjustbox{stack=cc, margin=0ex 2ex}{1s, no overlap} & \adjustbox{stack=cc, margin=0ex 2ex}{2022} & \adjustbox{stack=cc, margin=0ex 2ex}{\cite{peng2022domain}}  \\

        \rowcolor{Gray}
        \adjustbox{stack=cc, margin=0ex 2ex}{CNN (feature extraction),\\LSTM (classification)}  & \adjustbox{stack=cc, margin=0ex 2ex}{TUSZ}& \adjustbox{stack=cc, margin=0ex 2ex}{acc: 82\%,\\pre: 71.69\%,\\sen: 85\%}  & \adjustbox{stack=cc, margin=0ex 2ex}{LOO} & \adjustbox{stack=cc, margin=0ex 2ex}{N/A} & \adjustbox{stack=cc, margin=0ex 2ex}{2020} & \adjustbox{stack=cc, margin=0ex 2ex}{\cite{einizade2020deep}}  \\

        \adjustbox{stack=cc, margin=0ex 2ex}{Scalp Swarm Algorithm\\(SSA) (feature selection),\\LSTM (classification)} & \adjustbox{stack=cc, margin=0ex 2ex}{TUSZ}& \adjustbox{stack=cc, margin=0ex 2ex}{sen: 98.99\%,\\FDR: 98.43\%,\\spe: 99.01\%,\\acc: 99.2\%,\\F1: 97.54\%}  & \adjustbox{stack=cc, margin=0ex 2ex}{80-20\\train-test} & \adjustbox{stack=cc, margin=0ex 2ex}{1s, no overlap} & \adjustbox{stack=cc, margin=0ex 2ex}{2022} & \adjustbox{stack=cc, margin=0ex 2ex}{\cite{rani2022effective}}  \\

        \rowcolor{Gray}
        \adjustbox{stack=cc, margin=0ex 2ex}{RNN} & \adjustbox{stack=cc, margin=0ex 2ex}{CHB-MIT,\\TUEP}& \adjustbox{stack=cc, margin=0ex 2ex}{TUEP:\\acc: 84.7\%,\\sen: 89.2\%,\\spe: 82.2\%,\\CHBMIT:\\acc: 85.3\%,\\sen: 93.0\%,\\spe: 79.7\%}  & \adjustbox{stack=cc, margin=0ex 2ex}{90-10\\train-test} & \adjustbox{stack=cc, margin=0ex 2ex}{N/A}  & \adjustbox{stack=cc, margin=0ex 2ex}{2020} & \adjustbox{stack=cc, margin=0ex 2ex}{\cite{johnrose2021rag}}  \\

        \adjustbox{stack=cc, margin=0ex 2ex}{LSTM} & \adjustbox{stack=cc, margin=0ex 2ex}{CHB-MIT,\\Siena}& \adjustbox{stack=cc, margin=0ex 2ex}{Siena:\\acc: 92.59\%,\\sen: 94.83\%,\\spe: 96.82\%,\\CHBMIT:\\acc: 89.88\%,\\sen: 96.71\%,\\spe: 89.88\%}  & \adjustbox{stack=cc, margin=0ex 2ex}{10-fold\\CV} & \adjustbox{stack=cc, margin=0ex 2ex}{N/A} & \adjustbox{stack=cc, margin=0ex 2ex}{2019}  & \adjustbox{stack=cc, margin=0ex 2ex}{\cite{pandey2022epileptic}}  \\
        
        \aboverulesepcolor{Gray}
        \bottomrule
              
    \end{tabular}
    \end{adjustbox}
    \caption{continued - DL methods from systematic review of literature for seizure detection in scalp EEG data}
    \label{DL4}
\end{table} 

\newpage
\begin{table}[H]
\centering
    \begin{adjustbox}{width=\textwidth}
    \scriptsize	
    \begin{tabular}{>{\centering}p{0.25\textwidth}cccccp{0.1\textwidth}<{\centering}}
        \toprule 
        \textbf{Classifier}  & \textbf{Dataset(s)} & \textbf{Performance}  & \textbf{Validation} & \textbf{\makecell{Segment\\Length}} & \textbf{Year} & \textbf{References} \\
        \midrule
        \belowrulesepcolor{Gray}
        \rowcolor{Gray}

        \adjustbox{stack=cc, margin=0ex 2ex}{ConvLSTM} & \adjustbox{stack=cc, margin=0ex 2ex}{TUEP}& \adjustbox{stack=cc, margin=0ex 2ex}{acc: 92.17\%,\\sen: 93.27\%,\\spe: 90.96\%,\\pre: 91.23\%,\\F1: 0.93}  & \adjustbox{stack=cc, margin=0ex 2ex}{5-fold CV\\and LOO} & \adjustbox{stack=cc, margin=0ex 2ex}{3s,\\no overlap} & \adjustbox{stack=cc, margin=0ex 2ex}{2022} & \adjustbox{stack=cc, margin=0ex 2ex}{\cite{tawhid2022convolutional}}  \\

        \adjustbox{stack=cc, margin=0ex 2ex}{Convolution Attention Layer,\\ BiRNN classification} & \adjustbox{stack=cc, margin=0ex 2ex}{CHB MIT\\(patient 1-11, 14, 20-24)}& \adjustbox{stack=cc, margin=0ex 2ex}{acc: 97.62\%,\\sen: 96.69\%,\\spe: 98.41\%,\\F1: 97.38\%}  & \adjustbox{stack=cc, margin=0ex 2ex}{N/A} & \adjustbox{stack=cc, margin=0ex 2ex}{1s,\\no overlap} & \adjustbox{stack=cc, margin=0ex 2ex}{2022} & \adjustbox{stack=cc, margin=0ex 2ex}{\cite{xi2022two}}  \\
    
        \rowcolor{Gray}   
        \adjustbox{stack=cc, margin=0ex 2ex}{AE (feature extraction),\\RF (classification)} & \adjustbox{stack=cc, margin=0ex 2ex}{Siena}& \adjustbox{stack=cc, margin=0ex 2ex}{acc: 97.22\%}  & \adjustbox{stack=cc, margin=0ex 2ex}{LOO} & \adjustbox{stack=cc, margin=0ex 2ex}{6s,\\1s overlap} & \adjustbox{stack=cc, margin=0ex 2ex}{2022} & \adjustbox{stack=cc, margin=0ex 2ex}{\cite{ferariu2022detection}}  \\

        \adjustbox{stack=cc, margin=0ex 2ex}{CNN and CBAM (feature extraction),\\GRU (classification)} & \adjustbox{stack=cc, margin=0ex 2ex}{CHB-MIT\\(13 patients)}& \adjustbox{stack=cc, margin=0ex 2ex}{acc: 91.73\%,\\sen: 88.09\%,\\FPR:0.053/h\\spe: 92.09\%,\\AUC: 91.56\%}  & \adjustbox{stack=cc, margin=0ex 2ex}{10-fold\\CV} & \adjustbox{stack=cc, margin=0ex 2ex}{30s\\1s overlap} & \adjustbox{stack=cc, margin=0ex 2ex}{2023} & \adjustbox{stack=cc, margin=0ex 2ex}{\cite{ji2023effective}}  \\
        
        \rowcolor{Gray}   
        \adjustbox{stack=cc, margin=0ex 2ex}{2D-DCAE (feature extraction),\\Bi-LSTM (classification)} & \adjustbox{stack=cc, margin=0ex 2ex}{CHB-MIT\\(16 patients)}& \adjustbox{stack=cc, margin=0ex 2ex}{acc: 98.79 \(\pm\) 0.53\%,\\sen: 98.72 \(\pm\) 0.77\%,\\spe: 98.86 \(\pm\) 0.53\%,\\pre: 98.86 \(\pm\) 0.53\%,\\  F1: 98.79 \(\pm\) 0.53\%}  & \adjustbox{stack=cc, margin=0ex 2ex}{10-fold\\CV} & \adjustbox{stack=cc, margin=0ex 2ex}{4s,\\no overlap} & \adjustbox{stack=cc, margin=0ex 2ex}{2021} & \adjustbox{stack=cc, margin=0ex 2ex}{\cite{abdelhameed2021deep}}  \\

        %\aboverulesepcolor{Gray}
        \bottomrule
              
    \end{tabular}
    \end{adjustbox}
    \caption{continued - DL methods from systematic review of literature for seizure detection in scalp EEG data}
    \label{DL5}
\end{table} 

\subsection{Post-classification processing} \label{Post-classification processing}
We use the term post-classification processing to describe steps applied to the initial predictions a ML model has made on the input data. These steps include refining, improving, or interpreting the model's output before presenting it to the end-user. An important role for post-classification processing is to amalgamate classifier outputs in order to remove some outliers. This amalgamation can be across channels, in the case of algorithms trained on single-channel inputs applied to multi-channel data, and across time. This can be advantageous for increasing accuracy. However, it may increase latency when multiple segments are required to produce a final output.

It is generally accepted that in a clinical setting, high sensitivity and low false detection rate should take precedence over low latency. Indeed, devices producing false alarms can lead to a phenomenon known as `alarm fatigue', and are unlikely to be deployed in a clinical setting \cite{lewandowska2020impact}. Post-classification steps help in reducing the rate of false alarms, and are therefore important for the clinical translation of seizure detection algorithm. However, a minority of the encountered studies used any post-processing.

As mentioned above, post-processing steps for seizure detection algorithms can be divided into two categories: spatial and temporal. This is linked to the distinction between segment based and event based performance. 

\begin{enumerate}
    \item Spatial concatenation: Combining the predictions of different channels if the algorithm outputs a prediction per channel. For instance, only sending an alarm when 3 or more channels are classified as being in a seizure state within 5 seconds of each other.
    
    \item Temporal concatenation: Combining the predictions of consecutive EEG segments. This is applied by \citeauthor{khalkhali2021low}, who set \(BDmin\) and \(SDmin\) as a threshold for the minimum background and seizure duration, respectively \cite{khalkhali2021low}. All background events, in between seizure events, with a duration less than \(BDmin\) are converted to seizure events. All seizure events, in between background events, with a duration less than \(SDmin\) are converted to background events. Using this method, the maximum seizure detection delay is defined as: \(BDmin + SDmin\). \citeauthor{wei2020epileptic}, who arrived in fifth position at the 2020 Neureka\textsuperscript{TM} challenge (a seizure prediction challenge using the v.1.5.2 TUSZ dataset), grouped seizure labels if they were less than 2 seconds apart and classified an event as a seizure if it was longer than 15 seconds \cite{wei2020epileptic}. 
    
\end{enumerate}

\citeauthor{chatzichristos2020epileptic} won the 2020 Neureka\textsuperscript{TM} competition \cite{chatzichristos2020epileptic}. They merged consecutive events and discarded short events following three rules. Firstly, seizure events less than 30 seconds apart are merged together. Secondly, merged seizure events, for which the mean probability of being a seizure is less than 82\%, are rejected.  Finally, seizure events of duration less than 15 seconds are rejected. 

Most post-processing steps do not incorporate domain specific knowledge about seizures. For example, in patients on anti-seizure medication, the median length of a generalized tonic–clonic seizure is 79.5 seconds, whereas that of a focal tonic seizure is 15 seconds \cite{meritam2023duration}. In the case of \citeauthor{chatzichristos2020epileptic}, half of the tonic-clonic seizures would be missed. 

Embedding medical and contextual information within the post-processing design could improve performance. For example, in the case of a seizure detection device used in the ICU, post-processing should lower the rate of false positive and increase the sensitivity to status epilepticus (see Section \ref{NCSE_section}). However, when keeping a seizure diary, logging shorter seizures may be useful. Indeed, seizure types like generalised tonic seizures have a median duration of 8 seconds, and are important to record and monitor \cite{meritam2023duration}.  Overall, post-processing steps require tuning to work best in different clinical settings where different forms of seizures are expected, tolerated and monitored. 
\subsection{Metrics for performance evaluation}
Metrics such as accuracy, sensitivity and specificity are good bench marking tools, and allow objective comparisons of algorithms. However, to provide an assessment of the clinical usability of an algorithm, it is important to use a combination of metrics that provide information about the expected behaviour of the algorithm in a clinical setting such as the rate of false positives, which the majority of studies do not include.  

\begin{figure}[h]
    \centering
    \includegraphics[width=0.4\textwidth]{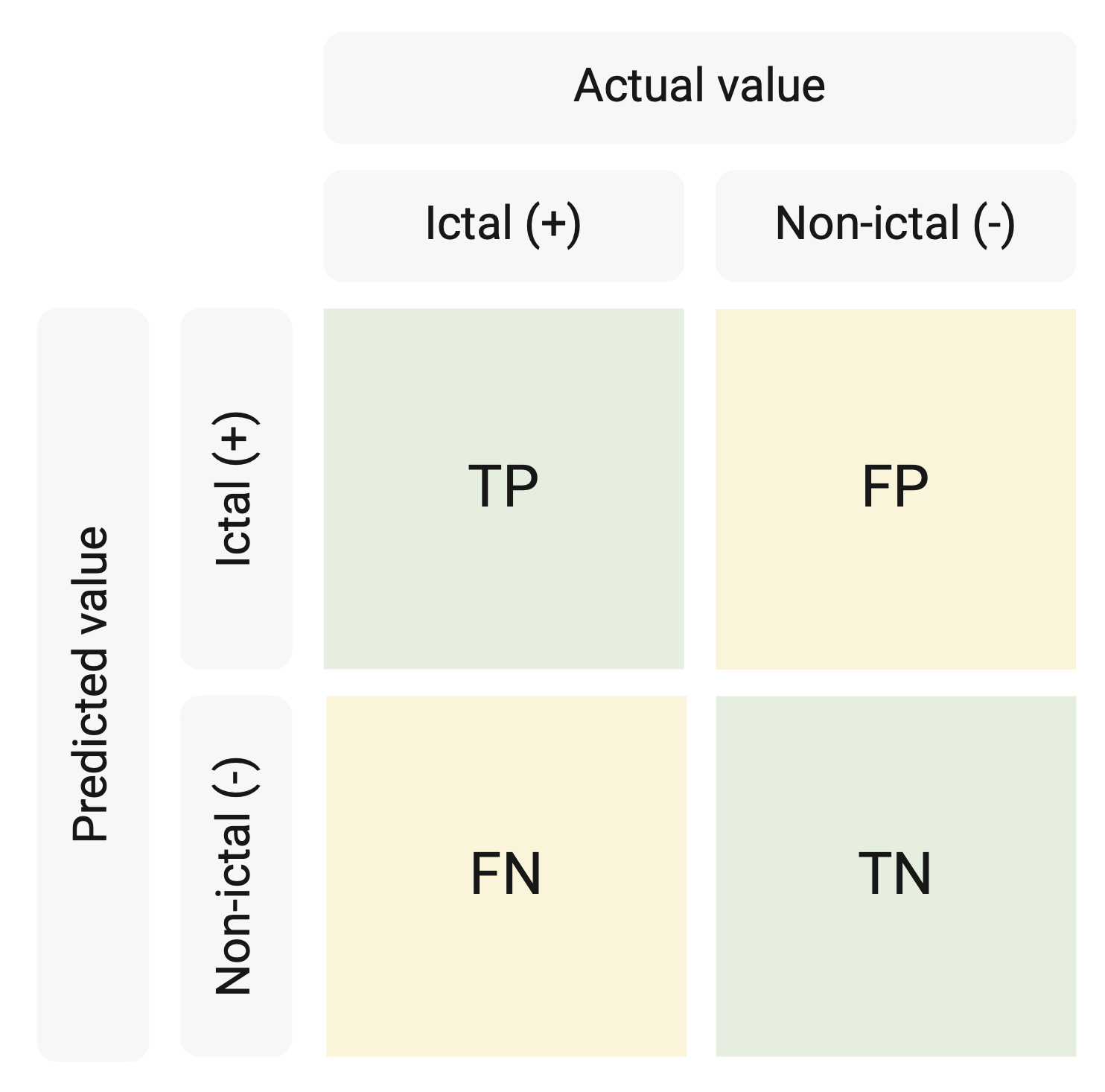}
    \caption{Confusion matrix of a binary seizure detection algorithm. Ictal segments are labeled as being positive and non-ictal segments as negative. TP = true positive, FP = false positive, FN = false negative, TN = true negative. (Created with BioRender, \cite{created_with_biorender})}
    \label{fig:confusionMatrix}
\end{figure}

For an exhaustive description of objective evaluation metrics in the field of automatic classification of EEG segments, see \cite{shah2021objective}. This section describes performance metrics and their clinical relevance with regards to seizure detection. The two-by-two confusion matrix in Figure \ref{fig:confusionMatrix} illustrates the binary classification results of seizure detection algorithms, from which several metrics are derived. Table \ref{tab:traditional_metrics} (Appendix) shows how accuracy, sensitivity, specificity, precision and F1 relate to the values of the confusion matrix.  

\subsubsection{Traditional evaluation metrics}

Accuracy is the most common metric for evaluating classification algorithms. It is defined as the proportion of correctly classified instances out of the total number of instances in the dataset. While accuracy is the most commonly used metric for evaluating the performance of classification models, it can be misleading. For example, if the test data is imbalanced and contains 1\% of ictal segments and 99\% of non-ictal segments, a model classifying all the instances as being non-ictal has an accuracy of 99\%. 

As discussed previously, EEG datasets used to test and train seizure detection algorithms are often imbalanced. Most studies balance the training data. However, this is not always the case for test data, where some authors try to mimic real likelihoods of ictal and non-ictal activity, i.e. reflect the true proportion of seizure/non-seizure activity, thereby using imbalanced data. In this case, other metrics than accuracy should also be reported. 

Even with balanced test data, the accuracy metric weights all classification errors (false positives, false negatives) equally, which does not necessarily reflect the real-world implications of different errors. In many applications of seizure detection algorithms, incorrectly classifying an ictal segment as non-ictal is worse than incorrectly classifying a non-ictal segment as an ictal segment. As such, it is necessary to use additional metrics, to assess  algorithm performance in a way that corresponds better to their clinical usability.
    
Sensitivity, also known as \textit{true positive rate} or \textit{recall}, is the proportion (or percentage) of positive instances that are correctly classified. A model with high sensitivity can effectively identify positive instances (ictal segments), i.e., true positives (TP). However, sensitivity is unrelated to how many negative instances are correctly labelled (TN). For example, if an algorithm assigns a positive label to all segments, it will have a sensitivity of 1. Sensitivity is independent to data imbalance as it only depends on the classification of the positive class, which is often the minority class.

Specificity, also known as \textit{selectivity} or \textit{true negative rate}, is complementary to sensitivity. Specificity is the proportion or percentage of the negative class (non-ictal) correctly predicted to be negative (TN).

Precision is the proportion of positive predictions that are true positives (TP). Unlike sensitivity and specificity, precision is affected by class imbalance as it considers the number of negative samples incorrectly labeled as positive.

The F1 score combines precision and recall. A high F1 score indicates that the model is good at identifying positive cases whilst avoiding false positives, while a low F1 score indicates that the model is not performing well in one or both of these areas. The formula for F1 score is given in Table \ref{tab:traditional_metrics} (Appendix).

The receiver operating characteristic curve (ROC curve) plots the true positive rate (sensitivity) over the false positive rate (1 - specificity) for different classification thresholds, depicting the trade-off between correctly classified positive samples and incorrectly classified negative samples. Some studies integrate the ROC curve to give the area under the receiver operating characteristic curve (AUC-ROC), which is a metric between 0 and 1. A value of 0.5 for a binary classifier indicates a performance no better than random guessing, and a value of 1 indicates perfect classification. However, some argue that AUC-ROC is misleading when dealing with imbalanced data \cite{davis2006relationship}. 

A more appropriate metric than the AUC-ROC when dealing with imbalanced datasets is the \textit{area under the precision recall (AUC-PR) curve}. The AUC-PR plots the precision over recall for different classification thresholds. This area is more representative of the algorithm's performance than AUC-ROC because precision is more sensitive to the presence of false positives than the FPR. In two papers from \citeauthor{salafian2021efficient}, the AUC-ROC is 13.79 \cite{salafian2021efficient} and 33.42 \cite{salafian2022cnn} percent higher than the AUC-PR, highlighting the importance of the metric choice on the apparent performance. Interestingly, both studies use the CHB-MIT dataset, but the ratio of ictal to non-ictal segments vary. In \cite{salafian2021efficient}, for every second of seizure data, there are 6 seconds of non-seizure data whereas in \cite{salafian2022cnn} for every second of seizure data, there are 20 seconds of non-seizure data. The study using the most imbalanced dataset (\cite{salafian2022cnn}) is also the one displaying the highest gap between the AUC-ROC and AUC-PR.

\subsubsection{Clinically-focused evaluation metrics}

The metrics mentioned above are not sufficient to assess the performance of an algorithm developed for clinical application. There is a growing need for a set of standardised scoring metrics that reflects the needs of end-users.

Latency is often used to evaluate the performance of seizure detection algorithms. Latency is the duration between the labelled start time of the seizure and the triggering of the seizure onset alarm. Latency is relevant when using online, or `live', seizure detection. Reducing latency enables quicker medical intervention, but often results in a higher rate of false positives – instances where the algorithm incorrectly signals a seizure onset. For instance, \citeauthor{khalkhali2021low}'s model has a detection latency of 300ms but a 42.05\% sensitivity and a rate of 5.78 false alarms per day \cite{khalkhali2021low}. Some studies point to a very low detection latency in support of their algorithm's efficacy. However, the associated disadvantages for accuracy may reduce an algorithm's clinical utility. Additionally, any ``arms race'' towards faster latency seizure detection should be tempered by the fact that it can be difficult to objectively define the onset of a seizure, with different experts labelling the start of the seizure at different times. Therefore, small reductions in latency, particularly if they compromise accuracy, do not necessarily represent a clinical advancement. Nevertheless, latency can be important for evaluating the clinical relevance of an algorithm and we encourage authors to provide this measure.

Another important metric for successful clinical deployment is the rate of false alarms (FAs) \cite{cvach2014managing}. A high rate of false alarms can desensitise medical staff, a phenomenon known as ``alarm fatigue'' which leads clinicians to ignore critical alarms, potentially endangering patients \cite{bach2018managing, bridi2014clinical}. Lower rates of false positives are needed for clinical deployment. The false positive rate, often expressed as per 24 hours, is one of the most important metrics and should always be reported.

Time-aligned event scoring (TAES) is a metric that was developed for the Neureka\textsuperscript{TM} epilepsy challenge. The metric combines sensitivity, the rate of false positives, and the proportion of channels used (out of 19 EEG channels). TAES is calculated as follows:
\[\textnormal{TAES} = \textnormal{Sens} - \alpha \times \textnormal{FAs}_{24\textnormal{hr}} - \beta \times N/19 \] 
where, \(\textnormal{Sens}\) is the sensitivity, \(\textnormal{FAs}_{24\textnormal{hr}}\) is the number of false positives every 24 hours, then \(\alpha\) and \(\beta\) are constants defined by the organising committee (set to 2.5 and 7.5 respectively). Incorporating an extra EEG channel may improve the algorithm's sensitivity and reduce the rate of false alarms, but is penalised for increased complexity. The higher the TAES, the better.

\newpage

\newpage
\section{Outstanding research questions}\label{question_section}

In this section, we highlight the key research avenues for clinical translation of automated seizure detection algorithms identified after reviewing the selected articles.

\subsection{Generalisability}

The main challenge in the field of seizure detection remains the generalisability of algorithms. A robust algorithm should be able to generalise across individuals; patient demographics; EEG montages, electrode number and placement; datasets; and seizure types. 

Whilst some research approaches imply that making an algorithm patient specific is necessary to achieve good performance, we do not find that this is reflected in the recent literature. For instance, \citeauthor{qiu2022lightseizurenet} developed LightSeizureNet (LSN), a deep learning model for seizure detection \cite{qiu2022lightseizurenet}. They compared the patient dependent and patient independent version of LSN. The patient specific model achieves 99.77\% accuracy, 97.11\% sensitivity, and 99.78\% specificity, with 113,800 parameters and 3.7 million multiply-accumulate operations (MACs, a computational complexity metric). In comparison, the patient independent model achieves 97.09\% accuracy, 96.49\% sensitivity, and 97.09\% specificity, with 198,300 parameters and 6.2 million MACs. 

A number of algorithms have been designed to remove inter-patient information. \citeauthor{zhang2020adversarial} proposed an adversarial network to decompose scalp EEG data into patient-related and seizure-related latent spaces \cite{zhang2020adversarial}, thereby reducing inter-patient contributions, allowing a patient independent analysis of seizure activity. Similarly, \citeauthor{nasiri2021generalizable} rely on adversarial network methods to learn patient invariant representations and reduce patient specific variations \cite{nasiri2021generalizable}. \citeauthor{jiang2019transfer} use transfer component analysis (TCA) to reduce the impact of individuals on EEG characteristics. TCA maps the original high-dimensional feature space to a lower dimensional subspace, where the data of individuals all follow the same distribution. The low-dimensional features are then input to a k-NN algorithm for classification \cite{jiang2019transfer}. \citeauthor{thuwajit2021eegwavenet} employ transfer learning to fine-tune a patient independent model to better fit a specific individual's data \cite{thuwajit2021eegwavenet}. By using approximately 1 hour of an individual's data, the models accuracy increases from 88.41\% ($\pm$15.23) to 96.69\% ($\pm$3.59) on that individual.

In summary, patient independent algorithms seem to reach performances similar to that of patient specific algorithms. It is also clear that the implementation of patient dependent algorithms is far more limited. Indeed, patient dependent algorithms can be useful for analysing long-term EEG recordings obtained using, for example, subcutaneous or intracranial EEG recordings. However, in a setting such as the ICU, algorithms should be patient independent as there is no time for training the model using the patient's data. Future research should focus on patient independent solutions, using a validation scheme mimicking a clinical environment such as the leave-one-out (LOO) validation strategy (see Section \ref{validation_section}). 

Although inter-patient generalisability is addressed in the literature, only one of the studies addresses generalisability across datasets. Typically, algorithms are trained and tested on a single dataset. This can lead to overfitting, for example, to the EEG hardware and recording technique used in a particular dataset. This possibility was confirmed by a study conducted by \citeauthor{yan2019automated} \cite{yan2019automated}. \citeauthor{yan2019automated} trained a CNN using spectrograms obtained from the CHB-MIT dataset and tested it on (1) a held out subset of the CHB-MIT dataset and, (2) private data, which they collected at the New York Presbyterian – Weill Cornell Medical Center (NYP-WC). For seizures visible on a spectrogram, they achieved sensitivity and specificity of >90\% using the CHB-MIT test set, and >90\% sensitivity but only 75–80\% specificity using the NYP-WC test set. The reduction in performance from changing the test dataset suggests that the model was overfitting to some of the CHB-MIT’s recording properties. We recommend training and testing algorithms on multiple datasets. 

In addition to generalisability across patients and datasets, algorithms should generalise to all seizure types. Despite the qualitative differences between seizure types clinically (e.g. focal seizures versus generalised seizures), and their differences in terms of standard EEG interpretation, the possibility of variability of accuracy across all seizure types is often overlooked in the seizure detection literature. Although not the main focus of this review, a few studies evaluated classification performance in relation to specific seizures types. For example, using a type of wavelet decomposition, fractal analysis, and a SVM for classification, \citeauthor{tang2020accurate} achieve an accuracy >95\% for all seven seizure subtypes in the TUSZ dataset \cite{tang2020accurate}. A clinically very important form of seizure is that of status epilepticus, a medical emergency (see Section \ref{NCSE_section}). None of the encountered public seizure datasets contain this form of seizure. Researchers should seek to include this type of seizures in their research. Given the scarcity of recordings of certain seizure types, using data augmentation techniques to increase the relative number of EEG recordings associated with rare types of seizures may be one approach to improve the performance of algorithms. 

\subsection{Variability in the ground truth labels}

There is subjectivity in the seizure labelling of scalp EEG done by clinicians, especially given the numerous artefacts it is corrupted by, which means that the ground truth labels used for training are subjective. In a 2009 study, \citeauthor{ronner2009inter} asked nine clinicians with different levels of experience to evaluate 90 epoch (10s each) of 30 EEG recordings of 23 different patients admitted to the ICU \cite{ronner2009inter}. For each epoch, clinicians had to decide whether there was an electrographic seizure or not. The results show a limited inter-observer agreement. The labelling of the more experienced clinicians obtained a Kappa value of 0.5 (moderate agreement) and that of the less experienced clinicians a Kappa value of 0.29 (fair agreement). Quantifying the labelling variance would require access to large amounts of labelled data from a wide range of clinicians, but could help develop more robust algorithms.

In April 2023, \citeauthor{jing2023interrater} asked 30 clinicians to  score varying numbers of ten-second EEG segments as seizures (SZ), generalized periodic discharges (GPDs), lateralized periodic discharges (LPDs), lateralized rhythmic delta activity (LRDA), generalized rhythmic delta activity (GRDA), or other \cite{jing2023interrater}. In total, clinicians scored 50,697 EEG segments. Results show that the average percent agreement with group consensus (i.e. majority voting) was 65\%. They suggest that the observed variations in labelling are due to variations in decision thresholds, rather than level of experience. In May 2023, they released what seemed to be a slightly modified version of this dataset as part of the harmful brain activity classification contest \cite{jing2023interrater, jing2023development}. The dataset contains 71,982 10 seconds long scalp EEG segments that have been independently annotated by 20 fellowship-trained neurophysiologists. This is the first dataset to enable quantification of labelling heterogeneity, which could be very useful to develop probabilistic classification algorithms. As of April 2024, to the best of our knowledge, it was retracted and has not yet been re-released as the authors are double-checking data quality. 

\subsection{Robustness to non-ictal activity} \label{ictal-interictal continuum}

An emerging challenge for automated seizure detection research is the \textit{ictal-interictal continuum (IIC),} which introduces a new dimension of complexity for algorithm development. The IIC covers a range of EEG signatures typically associated with critically ill patients. IIC patterns can be sharp, rhythmic or periodic and risk being incorrectly classified as ictal. For example, in critically ill patients one may encounter abnormal and often stereotyped bursts of striking epileptiform activity. If these bursts last less than 10 seconds, they are known as brief (potentially) ictal rhythmic discharges (BIRDs) and are not classified as seizures \cite{hirsch2021american}. Critically ill patients may often suffer from both frequent seizures (>10 seconds) with BIRDs (<10 seconds) in between. BIRDs are an intermediate phenomenon between short runs of standard interictal epileptiform discharges and definitive electrographic seizures. They are regarded as a marker of cortical irritability or hyperexcitability, and are on the IIC. It is important the seizure detection algorithms are robust to the presence of IIC patterns, to ensure that IIC segments are not misclassified as ictal activity  \cite{kamousi2021monitoring}. 

Finally, testing algorithms using scalp EEG recordings of healthy patients would be useful to assess their specificity. A number of publicly available datasets could be used for this purpose. For example, the Leipzig study for mind-body-emotion interactions (LEMON) dataset is composed of 227 healthy participants across a range of ages, making it a potentially suitable algorithmic control test set \cite{babayan2019mind}.

\subsection{Advancing epilepsy detection with SEEG} \label{SEEG-EEG integration}
Stereo electroencephalography (SEEG) is the surgical placement of electrodes deep within the brain to record electrical activity from the cerebral cortex directly and subcortical areas, offering precise spatial resolution and clear, localised seizure onset zones \cite{bartolomei2017defining}, as described in Section \ref{EEG_rec_techniques}. In contrast, scalp-EEG, a non-invasive method, records electrical activity through electrodes placed on the scalp, providing a more general and widespread view of brain activity with lower spatial resolution, as described in Section \ref{sEEG_section}. Simultaneously collecting SEEG and scalp-EEG signals presents a unique opportunity to enhance the understanding and detection of epileptic seizures. The clear onset of seizures detected by SEEG can be leveraged to refine and train scalp-EEG-based models for earlier and more accurate seizure detection.

However, using SEEG is not without challenges, primarily due to the high cost and invasive nature of the surgery required to implant the electrodes. To maximise the benefits of the simultaneous signals, data from both modalities can be integrated to enhance the localization of seizure sources. This integrated approach can lead to more precise mappings of epileptic networks, aiding in surgical planning and potentially reducing the extent of intervention required. Moreover, the simultaneous collection of SEEG and scalp-EEG data opens up avenues for advancing seizure prediction models. Researchers can enhance the sensitivity and specificity of non-invasive EEG models by analyzing how specific seizure signatures from SEEG correlate with broader EEG patterns.

\subsection{Virtual brain}

All the ML algorithms encountered in this review use a data-based approach for identifying ictal/non-ictal instances. In contrast, there is increasing interest in a ``bottom-up'' modelling approach to seizures, based on biophysical simulations of neural activity coupled to an EEG forward model. The Virtual Brain (TVB) is a publicly available toolkit for such mesoscopic simulations \cite{sanz2013virtual}. In the future, these models may provide valuable insights into the relationship between neural activity and observable EEG patterns. Reproducing observed EEG signals/features using these models could shed light on the underlying dynamics of seizure genesis and propagation.  The application of such models to individual patients is the basis of  the so-called ``digital twin'' approach, where the TVB has seen validation in the context of epilepsy surgery planning \cite{jirsa2017virtual, jirsa2023personalised}. Additionally, a bottom-up modeling approach could supplement limited empirical data by generating synthetic, artefact-free, EEG. 

\subsection{Wearable recording devices}

Traditional clinical EEG is costly and requires experts to set up and interpret.  Hence, EEG in most healthcare settings is not frequently used, even in ICUs, where seizure occurrence is high \cite{strein2019prevention}. Using several continuous EEG monitoring studies published between 1994 and 2011, \citeauthor{westover2015probability} established the occurrence of seizures in the ICU to be between 8-34\% \cite{westover2015probability}. As part of their study, \citeauthor{claassen2004detection} found that of 570 patients who had continuous EEG monitoring, 19\% (110) had seizures, 92\% of which were non-convulsive (102). Out of those 110 patients, 89\% (98) were in ICU at the time of monitoring \cite{claassen2004detection}. 

Technological advancements in wearable EEG devices is promising for addressing these issues and revolutionising EEG monitoring. One such portable device, the point-of-care EEG (POC-EEG) by Ceribell, consists of a headband with ten electrodes connected to a small battery powered recorder equipped with a screen for real-time EEG streaming (see \url{https://ceribell.com}). In a single centre cohort study, \citeauthor{rajshekar2023automated} found that in 72\% of patients monitored, POC-EEG was thought to have expedited diagnostic testing and/or treatment \cite{rajshekar2023automated}. Another alternative is in-ear or behind-the-ear EEG recording devices. The simple setup of ear-EEG devices may enable more widespread use of EEG in hospital and community settings, enabling simultaneous recording of EEG and ECG \cite{goverdovsky2017hearables, goverdovsky2015ear, nakamura2019hearables}. For this reason, algorithm design and evaluation for such devices becomes increasingly important. The ICASSP 2023 Seizure Detection Challenge is currently the only seizure detection competition using behind-the-ear EEG data \cite{al2023towards}.

\newpage
\section{Conclusion and algorithm development guidelines} \label{guidelines}
The worldwide economic burden associated with epilepsy in 2019 was estimated at \$119.27 billions \cite{begley2022global}. Whilst ML based algorithms have demonstrated significant potential for managing seizure disorders, there remain technological, clinical, and regulatory challenges for their successful translation.

In this review, we have provided a review of the key considerations for creating a robust, accurate, and clinically relevant EEG-based seizure detection system. Based on this analysis, we have developed the following recommendations for creating more clinically informed seizure detection algorithms.

\textbf{Identify the intended clinical use(s) of the algorithm.} Applications for seizure detection algorithms range widely, from (1) highlighting to clinicians sections of interest in long recordings to facilitate annotation offline; (2) real-time seizure detection using continuous EEG in ambulatory patients, telemetry units, or ICU; to (3) automatic recording of seizure diary entries. Identifying the specific intended use case(s) of the algorithm should be seen as a prerequisite for designing the algorithm. 

\textbf{Leverage multi-domain data representations.} To optimally distinguish between ictal and non-ictal EEG segments, researchers should leverage a range of EEG data representations, i.e. in the time domain as well as spectral and network domains. \citeauthor{yan2019automated} showed that there is no guarantee that all seizures are spectrographically visible (i.e. using spectral domain-based features), and subtle seizures, although visible on the EEG, may not be distinguishable from the spectrogram background. As such, we recommend using multiple domain representations as it is likely to improve performance. A number of high throughput feature extraction libraries support multi-domain data representations. For example, the MATLAB package \emph{hctsa} extracts 7,749 different features from time-series data \cite{fulcher2013highly}. This approach can help identify discriminating features for a given classification task, and provides an exhaustive overview of a signal's properties. When using feature-based ML techniques, we recommend using similar tools during the feature selection process, or providing the rationale behind the choice of features.

\textbf{Use relevant performance metrics.} The performance metrics encountered in the literature are varied and reflect different aspects of algorithmic performance. We recommend providing all metrics in Table \ref{tab:traditional_metrics}, latency, and the rate of false positives. 

\textbf{Explore post-classification processing.} The input rate of EEG segments into an algorithm is typically less than 5 seconds. In post-processing, the use of multiple successive predictions increases the algorithm's ability to correctly classify ictal/non-ictal activity. We suggest evaluating the effects of post-classification processing, and using multiple consecutive EEG segments during classification. We also encourage authors to provide the rationale behind their choice of post-processing steps, as they may be influenced by the use-case of an algorithm.

\textbf{Use diverse training data.} We recommend training and testing algorithms on a range of datasets. This approach enables greater understanding of which features and/or architectures are most independent of patients, seizure types and hardware recording devices. Finally, the results could be validated using additional seizure-free control datasets like the LEMON dataset (healthy participants) \cite{babayan2019mind}. For generalised medical applications, using multiple datasets reduces sensitivity to recording equipment, cohort community, and provides a larger dataset.

\textbf{Use a leave-one-out validation strategy.} We recommend using a leave-one-patient-out cross-validation strategy as it simulates real-world scenarios effectively. Excluding the data of one patient from the training set and using cross-validation evaluates the ability of the model to generalise across the entire patient cohort. Using a random train/test split may introduce sampling bias and overfitting.   

\textbf{Respect the assumption of independence.} A common pitfall in ML is the violation of the assumption of independence. The assumption of independence states that the data used for model training and evaluation is independent and identically distributed. This is a necessary condition to achieve unbiased estimates of performance metrics and generalisation error of the model. To avoid this common pitfall, oversampling, feature selection, and data augmentation should always be performed after the data was split into training and evaluation sets \cite{maleki2022generalizability}. 

\textbf{Model architecture and pre-processing steps.} Regarding the choice of model architecture, segment length, channels to use and other pre-processing steps; it is challenging to identify an optimal strategy. In fact, there is likely no `one size fits all' strategy, as the use case of the algorithm influences design choices drastically.

However, we found that segment lengths of 3-5 seconds (with a sampling rate >170 Hz) were good candidates for seizure detection as they are quasi-stationary and seem to carry sufficient information. We recommend using a notch filter to remove the power line interference and a band-pass filter between 0.3-60 Hz. For channel selection, we recommend using all of the available channels for seizure detection. Depending on the type of seizure and its location, any of the 10-20 electrodes can pick up seizure activity. Given the absence of a subset of channels that is independent of patient and seizure types, and that also contains all necessary information to replicate detection results achieved with the complete set of channels, we recommend against discarding any channel data.

Both feature-based ML and DL approaches can achieve high seizure detection performances. We found that the average detection accuracy using feature-based ML was 94.3\% (95\% confidence interval: 94.3 ± 2.3), compared to 91.8\% (95\% confidence interval: 91.8 ± 2.3) using deep learning architectures. An interesting approach, highlighted in the work of \citeauthor{zhao2023epileptic} is the use of interpretable ML models to understand what part of the EEG recording a model `identifies' as seizure activity, and quantify how `confident' the model is \cite{zhao2023epileptic}.    

Finally, we highly recommend tailoring the algorithm to the intended clinical use. Whilst there is likely no `one size fits all' solution, we recommend providing justification behind the choice of features, pre-processing steps, model architecture, validation strategy, evaluation metrics and dataset(s).  

\newpage
\section*{Acknowledgments}
NM was supported by the UKRI Centre for Doctoral Training in Artificial Intelligence for Healthcare (EP/S023283/1).

\section*{Conflict of interest}
The authors declare no conflict of interest, financial or otherwise.

\newpage
%\section*{List of abbreviations}
%\newpage
%\printglossary[type=\acronymtype]

\section*{Appendix}

\begin{table}[h]
    \centering
    \begin{adjustbox}{width=1\textwidth}
    \small
    \begin{tabular}{ccc}
        \toprule 
        \textbf{Databases} & \textbf{Search string} & \textbf{Number of papers} \\
        \midrule
        \belowrulesepcolor{Gray}
        \rowcolor{Gray}
   
        \adjustbox{stack=cc, margin=0ex 2ex}{Scopus} & 
        \adjustbox{stack=cc, margin=0ex 2ex}{( TITLE-ABS-KEY ( eeg AND detection AND epilepsy OR seizure ) AND NOT TITLE-ABS-KEY ( animal OR neonatal OR infant OR intracranial OR ieeg ) ) AND PUBYEAR > 2018 AND PUBYEAR < 2024 AND ( LIMIT-TO ( DOCTYPE , "cr" ) OR LIMIT-TO ( DOCTYPE , "cp" ) OR LIMIT-TO ( DOCTYPE , "ar" ) ) AND ( LIMIT-TO ( LANGUAGE , "English" ) )} & 
        \adjustbox{stack=cc, margin=0ex 2ex}{1886} \\

        \adjustbox{stack=cc, margin=0ex 2ex}{PubMed} & 
        \adjustbox{stack=cc, margin=0ex 2ex}{((eeg AND detection) AND (epilepsy OR seizure) NOT (animal OR neonatal OR infant OR intracranial OR iEEG) AND (("2019"[Date - Publication] : "2023-02-22") AND (english[Language])} & 
        \adjustbox{stack=cc, margin=0ex 2ex}{1078} \\

        \rowcolor{Gray}
        \adjustbox{stack=cc, margin=0ex 2ex}{WoS} & 
        \adjustbox{stack=cc, margin=0ex 2ex}{(EEG detection) AND (epilepsy OR seizure) NOT (animal OR neonat* OR infant OR intracranial OR iEEG) (Topic) and Article OR Book Chapter OR Book OR Proceedings Paper (Document Type) and English (Language)} & 
        \adjustbox{stack=cc, margin=0ex 2ex}{1551} \\

        \adjustbox{stack=cc, margin=0ex 2ex}{ \textbf{TOTAL}} &  & 
        \adjustbox{stack=cc, margin=0ex 2ex}{\textbf{4515}} \\
        
        \bottomrule
    \end{tabular}
    \end{adjustbox}
    \caption{Search strategy for each repository. Each search string corresponds to equivalent inclusion criteria. They are formatted to fit the database specific search tool's syntax.}
    \label{prisma_search_string}
   
\end{table}

\begin{table}[h!]
    \tiny
    \footnotesize
    \begin{tabular}{ScScSc}
        \toprule 
        \textbf{Name} & \textbf{Formula} & \textbf{Description} \\
        \midrule
        \belowrulesepcolor{Gray}
        \rowcolor{Gray}
        
        Accuracy &  $\begin{aligned} \frac{\text{TP + TN}}{\text{TP + TN + FN + FP}} \end{aligned}$ & \makecell{Proportion of correctly classified instances or correct predictions out\\of the total number of instances in the dataset} \\
        
        \makecell{Sensitivity, \\ Recall} & $\begin{aligned}\frac{\text{TP}}{\text{TP + FN}}\end{aligned}$ & \makecell{Evaluate how well a classification algorithms\\identifies positive instances} \\
        \rowcolor{Gray}

        Specificity & $\begin{aligned}
        \frac{\text{TN}}{\text{TN + FP}} \end{aligned}$ & \makecell{Evaluate how well a classification algorithms\\identifies negative instances} \\

        Precision & $\begin{aligned} 
        \frac{\text{TP}}{\text{TP + FP}} \end{aligned}$ &  \makecell{Proportion of positive predictions that are actually true positives} \\ 
        \rowcolor{Gray}
   
        F1 & $\begin{aligned}
        \frac{\text{2 * Precision * Recall}}{\text{Precision + Recall}} \end{aligned}$ &  \makecell{It combines both precision and recall to provide a\\balanced view of a model’s performance} \\

        \aboverulesepcolor{Gray}
        \bottomrule
    \end{tabular}
    \caption{Table of common metrics used to evaluate the performance of ML algorithms. Note: sensitivity is also known as \textit{True Positive Rate} and specificity as \textit{True Negative Rate}}
    \label{tab:traditional_metrics}
    %\vspace*{1 cm}
\end{table}

\newpage

\printbibliography %Prints bibliography
\end{document}